\documentclass[journal=jpccck,manuscript=article]{achemso}
\SectionNumbersOn

\usepackage[version=3]{mhchem} 
\usepackage{color}
\usepackage{graphics}
\usepackage{amsmath}
\usepackage{nccmath}
\usepackage{epsfig}
\usepackage{fancyhdr}
\usepackage{float}
\usepackage[normalem]{ulem}
\usepackage{setspace}
\usepackage{wrapfig}  
\usepackage{soul}
\usepackage{amsmath,amssymb}
\usepackage{tabularx}
\usepackage{svg}
\usepackage[utf8]{inputenc}

\usepackage{url} 
\usepackage{multicol}
\usepackage{multirow}
\usepackage[utf8]{inputenc}
\usepackage[T1]{fontenc}
\usepackage{graphicx}
\usepackage{textcomp}
\DeclareUnicodeCharacter{00E4}{\"a}

\author{Hassan A. Qureshi}
\author{Henri Lyyra}
\author{Akseli Korkeamäki}
\author{Oskar Tuomi}
\affiliation[Materials UTU]
{Department of Mechanical and Materials Engineering, University of Turku, FI-20014 Turku, Finland}
\author{Antti J. Moilanen}
\affiliation[EasternFInalnd]
{Center for Photonics Sciences, Department of Physics and Mathematics, University of Eastern Finland, FI-80100 Joensuu, Finland}
\author{Konstantinos S. Daskalakis}
\email{konstantinos.daskalakis@utu.fi}
\affiliation[Materials UTU]
{Department of Mechanical and Materials Engineering, University of Turku, FI-20014 Turku, Finland}

\title{A fully solution-processed organic microcavity laser in the strong light–matter coupling regime}

\date{31 October 2025}

\begin{document}

\section{Abstract}
Solid-state semiconductor lasers underpin technologies from telecommunications and data storage to sensing, medical diagnostics, and emerging quantum communication. Polaritons—hybrid exciton–photon states—have further extended this reach, enabling room-tem\-per\-a\-ture quantum effects such as low-threshold lasing and single-photon nonlinearities. Organic semiconductors are ideal for polaritonics due to their large exciton binding energy, strong optical nonlinearities, and straightforward processing, making them attractive for both classical and quantum photonics. While solution-processed organic films have been widely explored, their optical cavities have almost always been fabricated using vacuum deposition, limiting the realization of truly scalable and low-cost devices. Here, we report the first organic laser microcavities fabricated entirely by solution processing, which operate in the strong coupling regimeThe resulting platform can be driven reliably to high excitation densities, where we observe a reversible, interaction-driven redistribution of the polariton condensate, revealing a distinct polariton lasing behaviour in organic microcavities. Together, the fabrication approach and the observed lasing dynamics establish a route toward scalable polaritonic and quantum photonic technologies and provide new opportunities for studying nonlinear polariton physics in organic systems.

\section{Introduction}
Solid-state semiconductor lasers have become indispensable to modern technology. They form the backbone of global telecommunications and optical data storage, power everyday consumer devices, play a central role in medical diagnostics, and drive advanced sensing applications. Their efficiency, scalability, and spectral versatility further extend their impact into emerging frontiers such as quantum communication and information processing. As laser technologies continue to evolve toward integrated, low-cost, and quantum-capable platforms, new material systems and physical mechanisms are being explored\cite{Yoshida2023}. In particular, exciton-polaritons (polaritons hereafter) offer an alternative lasing paradigm that combines strong light–matter coupling with the potential for ultralow thresholds and rich quantum optical behavior. Unlike conventional photon-based lasing, polaritonlasing (often also referred to as polariton condensation) can occur without population inversion and may be achieved in systems where traditional lasing is challenging\cite{Deng2003}.

Polaritons have thus emerged as a powerful bridge between classical and quantum photonic technologies, enabling demonstrations from low-threshold lasing to single-photon nonlinearities\cite{Daskalakis2014, Plumhof2013,  Lerario2017, Rajendran2019, Zasedatelev2021, Tang2021, Moilanen2021, jiang2022exciton}. These hybrid exciton–photon states arise in the strong light–matter coupling regime, where the exchange of energy between excitons and cavity photons exceeds their individual dissipation rates\cite{Kavokin2011,torma_strong_2015,FriskKockum2019}. Organic semiconductors are particularly well suited to support polaritons at room temperature, owing to their large exciton binding energy. Moreover, they are attractive as solid-state laser materials due to their strong optical nonlinearities and straightforward processing. Together, these features have driven intensive research into organic lasers, both as scalable platforms for classical photonics and as versatile building blocks for emerging quantum devices. 

\begin{figure}
\centering
\includegraphics[width=\linewidth]{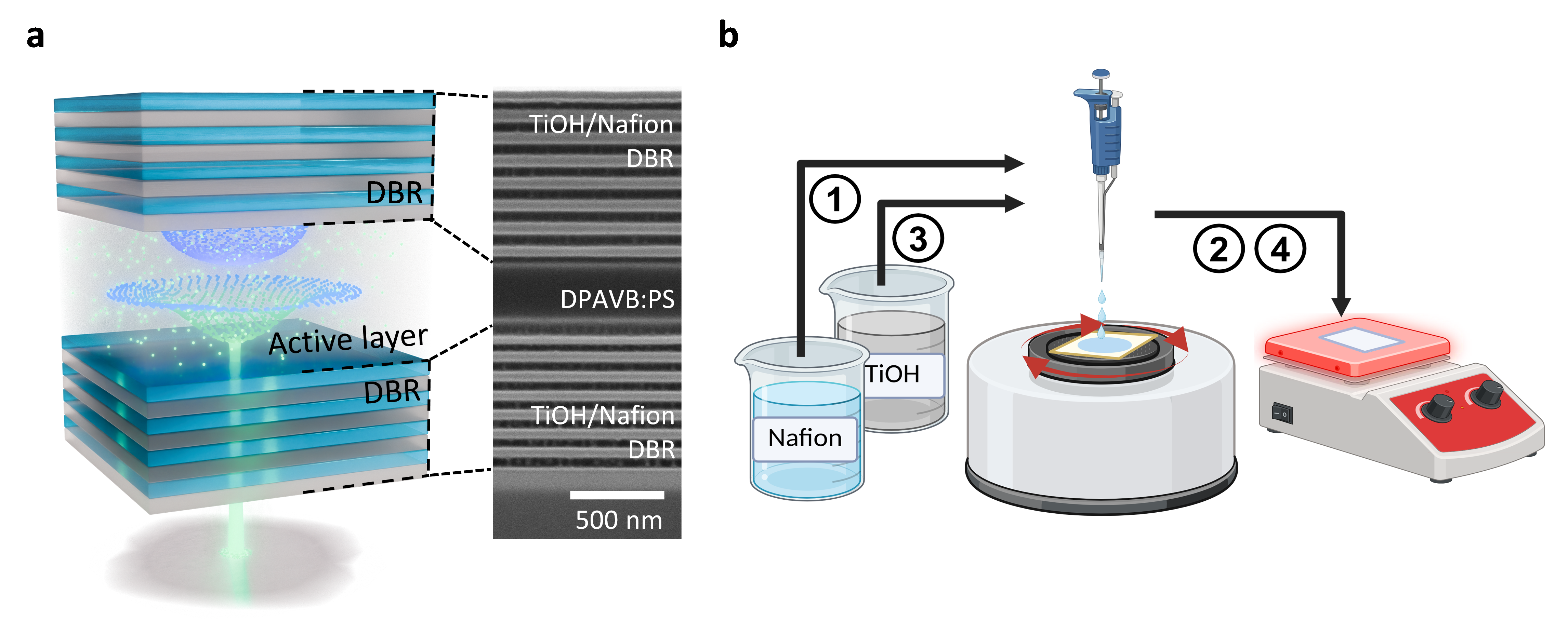}
\vspace{-20pt}
\caption{(a) Schematic illustration and corresponding scanning electron microscopy image of the all-solution-processed microcavity composed of alternating Nafion and titanium hydroxide/poly(vinyl alcohol) hybrid films as distributed Bragg reflectors (DBRs), with a 260-nm-thick DPAVB:polystyrene film as the active layer. (b) Schematic of the spin-coating process used for DBR fabrication. Each DBR pair is deposited through the following steps: (1) spin coating of titanium hydroxide/poly(vinyl alcohol) solution, (2) annealing, (3) spin-coating of Nafion, and (4) annealing.}
\label{fig:1}
\vspace{0pt}
\end{figure}

A natural next step is to leverage the inherent solution processability of organics to realize fully solution-processed lasers—offering a low-cost, scalable alternative to conventional deposition-based fabrication. While numerous studies have demonstrated lasing from solution-processed organic films, the cavities in these systems are almost invariably fabricated using vacuum-based deposition, undermining the promise of a fully solution-based approach\cite{palo2023}. Overcoming this limitation is crucial for advancing organic lasers toward practical, cost-effective technologies. Building on our previous work on fully solution-processed polariton microcavities\cite{Qureshi2025,palo2023developing}, we now demonstrate polariton lasing from microcavities fabricated entirely by spin coating (Fig.~\ref{fig:1}). In this work, we classify the observed emission as polariton lasing based on the combined observation of a nonlinear threshold, linewidth narrowing, energy blueshift without convergence to the bare cavity mode, and emission following the lower polariton dispersion, indicating that strong coupling is preserved above threshold (Fig.~\ref{fig:3}).
To our knowledge, no conventional optically pumped organic vertical-cavity lasers have yet been realized using solution-based methods. This suggests that strong light–matter coupling may be a key enabler of lasing in fully solution-processed vertical microcavities. Above threshold, the condensate also undergoes a spatial and momentum redistribution driven by repulsive polariton–polariton and polariton–reservoir interactions, reflecting the interacting nature of the condensate. In addition, our analysis shows that the condensate exhibits a thermal distribution, with effective temperature decreasing toward room temperature at high polariton densities, further underscoring the potential of this platform for scalable polaritonic devices in both classical and quantum photonic applications.

\section{Results}

The microcavities investigated here were fabricated entirely by spin coating, with both the dielectric mirrors and the lasing active medium deposited using this method. Their architecture and fabrication steps are schematically illustrated in Figs.~\ref{fig:1}a and \ref{fig:1}b, respectively. The active layer consists of the organic molecule 4-(Di-p-tolylamino)-4'-[(di-p-tolylamino)styryl]stilbene (DPAVB) embedded in an optically inert polystyrene (PS) matrix, forming a single film with a tunable thickness of ~260–270 nm, controlled by spin coating conditions (see Methods). The experimentally
measured complex refractive index of the active layer is shown in Supplementary Fig.~\ref{S_fig:Transfer matrix}c. The cavity mirrors are dielectric distributed Bragg reflectors (DBRs) composed of alternating quarter-wave TiOH:PVA and Nafion layers. While these DBRs were initially developed in our earlier work~\cite{palo2023developing,Qureshi2025} using dip coating, adapted from
\citeauthor{Bachevillier2019}\cite{Bachevillier2019}, here we transitioned to spin coating to improve film uniformity. This approach produces mirrors with high reflectivity ($\sim 94\%$, Supplementary Fig.~\ref{S_fig:DBR_ref}) and excellent interfacial integrity (right side of Fig.~\ref{fig:1}a), resulting in optical cavities with a quality factor \textcolor{red}(Q) > 325 for the 7.5-pair DBRs used in the lasing devices—among the highest reported for planar polariton microcavities~\cite{Bhuyan2023}.

\begin{figure}
\vspace{-5pt}
\centering
\includegraphics[width=0.99\linewidth]{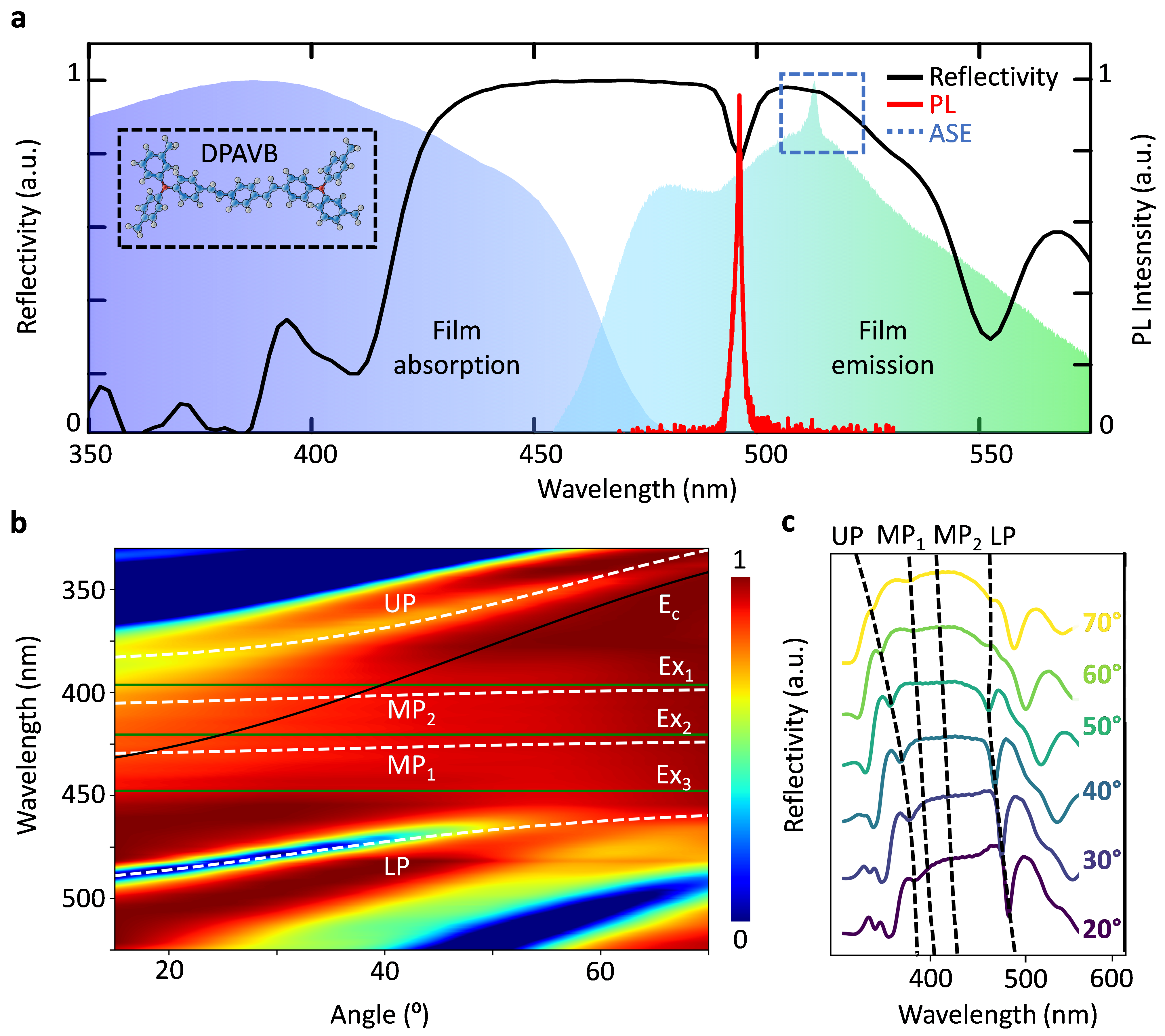}
\vspace{-12pt}
\caption{ (a) Reflectivity spectrum of the polariton microcavity at normal incidence (black line), overlaid with the absorption (blue shaded) and emission (green shaded) spectra of the active material. The red peak corresponds to the photoluminescence (PL) from the lower polariton band bottom. The black dashed box shows the DPAVB molecule while the blue dashed box highlights the ASE region}. (b) Angle-resolved reflectivity map of the microcavity with coupled oscillator model fit, showing the UP, cavity mode (E$_c$), exciton resonances (Ex$_1$, Ex$_2$, Ex$_3$), middle polaritons (MP$_1$, MP$_2$) and LP. (c) Corresponding waterfall plot of angle-resolved reflectivity spectra from (b) with polariton branches indicated. To enhance the visibility of the polariton modes, the measurements shown in panels (b) and (c) were performed on microcavities with a top DBR consisting of 5 pairs, instead of the 7.5 pairs used elsewhere in the manuscript.
\label{fig:2}
\vspace{-5pt}
\end{figure}

DPAVB is a commercially available fluorescent dopant dye that has been used in blue OLEDs, for amplified spontaneous emission (ASE), and—most relevant to this work—to demonstrate polariton lasing using purpose-optimized physical vapor deposition (PVD)\cite{mcghee_polariton_2022}. A key reason for selecting DPAVB was its good solubility in non-polar solvents such as toluene, which makes it compatible with non-polar polymer hosts like polystyrene. We therefore used polystyrene as the matrix material to prepare the emissive layer. This choice was crucial because our DBR structure contains PVA and Nafion, both of which are soluble in polar solvents. 
Figure~\ref{fig:2}a shows the normal-incidence reflectivity of the polariton microcavity together with the absorption, photoluminescence, and ASE spectra of DPAVB in a PS matrix. ASE is observed at 513~nm with a threshold pump fluence of 7~µJ/cm$^{2}$ (Supplementary Fig.~\ref{S_fig:ASE_PD}). For comparison, we also include the polariton photoluminescence spectrum at normal incidence, confirming the spectral overlap between the cavity mode and the molecular emission. 

Figures \ref{fig:2}b,c display the angle-dependent reflectivity of a 270 nm-thick microcavity as a color map. For these measurements, the top DBR consisted of 5 pairs—reduced from the lasing devices to improve mode visibility. Clear anticrossings between the cavity photon mode and three excitonic resonances (Ex$_1$, Ex$_2$, Ex$_3$) are observed, yielding four branches (UP, MP$_2$, MP$_1$, and LP). 
Fitting these dispersions with a coupled-oscillator model yields Rabi splittings of $\approx$~0.230~eV, 0.160~eV, and 0.130~eV for the respective exciton–photon interactions strength, matching previous reported values of Rabi splitting in DPAVB\cite{mcghee_polariton_2022}. Note that our cavity linewidth is <~10~meV thus satisfying the strong coupling condition of $\hbar \Omega_R > \frac{1}{2}$ (cavity linewidth $+$ exciton linewidth). The observed exciton–photon coupling in our samples is among the highest reported in solution processed microcavities~\cite{palo2023developing,Strang2024,Qureshi2025}. Transfer-matrix method (TMM) reflectivity simulations for the same cavity design (Supplementary Fig.~\ref{S_fig:Transfer matrix}a) show excellent agreement with the experimental data of Fig.~\ref{fig:2}b. Comparison with the empty cavity TMM reflectivity simulation highlights how the anti-crossing emerges when the emitters are introduced to the cavity (Supplementary Fig.~\ref{S_fig:Transfer matrix}b).

\begin{figure}
\vspace{-5pt}
\centering
\includegraphics[width=\linewidth]{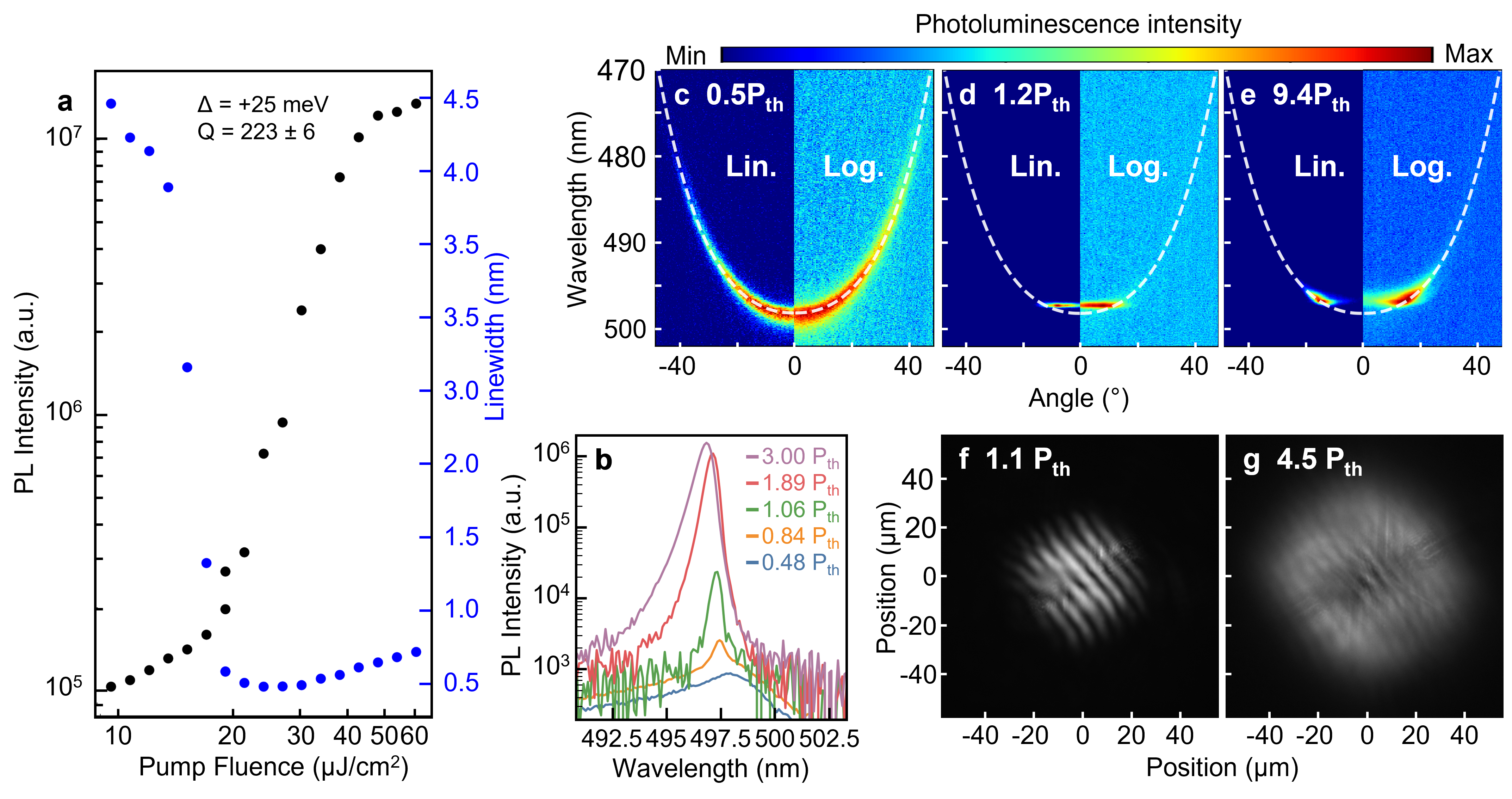}
\vspace{-12pt}
\caption{ (a) Power-dependent LP photoluminescence intensity (black dots) and linewidth (blue dots). $\Delta$ is the exciton-cavity detuning and Q is determined from five low-fluence measurements integrated over small collection angles. (b) Representative LP photoluminescence spectra recorded below and above the lasing threshold. (c–e) Angle-resolved LP emission below threshold (c) and above threshold (d, e). A sharp collapse of the emission into the bottom of the LP dispersion is observed at threshold (d), followed by a progressive blueshift and redistribution at higher excitation densities (e). The white dashed line in c-e) is a polynomial fit to the LP dispersion at 0.5~$P_{\mathrm{th}}$, showing that the condensate remains confined within the below-threshold LP branch. The left-hand side of panels (c–e) is plotted on a linear scale to highlight qualitative spectral evolution, while the right-hand side uses a logarithmic scale to visualize how well the condensate remains confined within the LP dispersion. (f, g) Michelson interferograms of the condensate at 1.1~$P_{\mathrm{th}}$ and 4.5~$P_{\mathrm{th}}$, respectively. (f) Just above threshold, clear interference fringes emerge, indicating the buildup of spatiotemporal coherence. (g) At higher excitation densities, the fringes extend over the entire excitation area but become less distinct.}
\label{fig:3}
\vspace{-5pt}
\end{figure}

Figure~\ref{fig:3} summarizes the power-dependent emission response of the polariton microcavities under nonresonant optical excitation at room temperature, with the notable feature of a polariton lasing threshold at 20~µJ/cm$^{2}$. Below threshold, the photoluminescence follows the LP dispersion (Fig.~\ref{fig:3}c), and its intensity (black dots in Fig.~\ref{fig:3}a) increases linearly with the absorbed pump fluence, indicating the absence of significant bimolecular annihilation. At threshold, a pronounced nonlinear increase in emission is accompanied by substantial spectral narrowing (blue dots in Fig.~\ref{fig:3}a) and the LP linewidth (full width at half maximum, FWHM) is reduced from above 4.5~nm 22.5~meV) to below 0.5~nm (2.5~meV). The corresponding power-dependent LP photoluminescence spectra, collected within $\pm 48$ deg, is shown in Fig.~\ref{fig:3}b. Comparison between similar PL power series measurements with different combinations of excitation and collection polarizations indicates negligible
cavity anisotropy or birefringence (Supplementary Fig.~\ref{S_fig:more_lasing_info}). A Michelson interferometer in retroreflector configuration reveals clear interference fringes above the lasing threshold in Fig.~\ref{fig:3}f and \ref{fig:3}g, confirming the build-up of spatial and temporal coherence. Time-resolved photoluminescence measurements provide further evidence of polariton lasing, showing a sharp reduction in emission lifetime above threshold (Supplementary Fig.~\ref{S_fig:transient}).

Upon further increasing the excitation density, the emission undergoes a blue-shift accompanied by a shift to higher momentum. Unlike previous reports~\cite{yagafarov_mechanisms_2020,Daskalakis2014}, the condensate in our system remains confined within the below-threshold polariton dispersion rather than blue-shifting toward the bare cavity mode (signifying coupling-strength depletion through phase-space filling~\cite{arnardottir_multimode_2020,moilanen_mode_2022}) or linearizing at the exciton emission (cavity depletion). This behaviour is illustrated in Figs.~\ref{fig:3}c–e, where we plot the angle-resolved emission below threshold (0.5$P_{\mathrm{th}}$), just above threshold (1.2$P_{\mathrm{th}}$), and at higher excitation (9$P_{\mathrm{th}}$), using both linear (left panel) and logarithmic (right panel) colour scales. At 1.2$P_{\mathrm{th}}$, the emission collapses into the bottom of the LP band ($\Delta\theta \approx$~ $11^\circ$, Fig.~\ref{fig:3}d), accompanied by an emission peak blue-shift of 1.52~meV that nevertheless remains within the 10~meV LP mode linewidth below threshold. At 9.4$P_{\mathrm{th}}$, the energy blueshift of 3.55~meV moves the condensate to higher energies along the dispersion band, allowing the emission to redistribute away from the band bottom resulting in a momentum shift of 0.48 $\mu$m$^{-1}$. As a result, the condensate develops symmetric, mirrored components at finite angles (in-plane momentum), producing a distinctive “Dali's mustache-like” pattern (Fig.~\ref{fig:3}e).  In this regime, coherence persists and the interference fringes expand to cover the entire excitation spot but become less distinct, see Fig.~\ref{fig:3}g. Further analysis presented in Supplementary Figs.~\ref{S_fig:lasing_reflectivity} and ~\ref{S_fig:dispersion_detunings} confirms that the sample remains in the strong-coupling regime. Across the full excitation range, the emission stays confined within the lower polariton dispersion and does not evolve toward the bare cavity mode. Moreover, cavities with more photonic detuning exhibit larger energy blueshifts for comparable momentum shifts, consistent with the detuning-dependent curvature of the lower polariton branch rather than cavity-mode behaviour. These observations suggest interaction-driven renormalization within the strong-coupling regime.

It is worth noting that in earlier reports of DPAVB-filled microcavities, normal-incidence lasing was observed across multiple modes at the polariton band bottom, but only when the polariton was tuned to 496~nm~\cite{mcghee_polariton_2022}. By contrast, in our devices such multimode behaviour appears only when the emission is collected without a polarizer. Moreover, as shown in Supplementary Fig.~\ref{S_fig:Lasing_in_other_det}, lasing can be achieved at different detunings, with thresholds governed by the cavity quality factor rather than detuning. Due to the inherent dynamics of spin coating, the cavity exhibits a thickness gradient across the substrate, resulting in spatially varying detuning (Supplementary Fig.~\ref{S_fig:polariton_photo}). By translating the optical collection area, multiple detunings can therefore be accessed on a single sample. The measurements shown in Supplementary Figs. \ref{S_fig:Lasing_in_other_det}, \ref{S_fig:dispersion_detunings}, and \ref{S_fig:ring_detunings} were performed at different spatial positions on two separate samples, corresponding to distinct cavity detunings. Supplementary Fig.~\ref{S_fig:Lasing_in_other_det} presents measurements obtained from two samples, with two detuning conditions investigated on each sample. Supplementary Figs. \ref{S_fig:dispersion_detunings} and \ref{S_fig:ring_detunings} show measurements performed on the higher-Q sample from this set, where three different detuning conditions were explored across different spatial regions.

\begin{figure}
\vspace{-5pt}
\centering
\includegraphics[width=\linewidth]{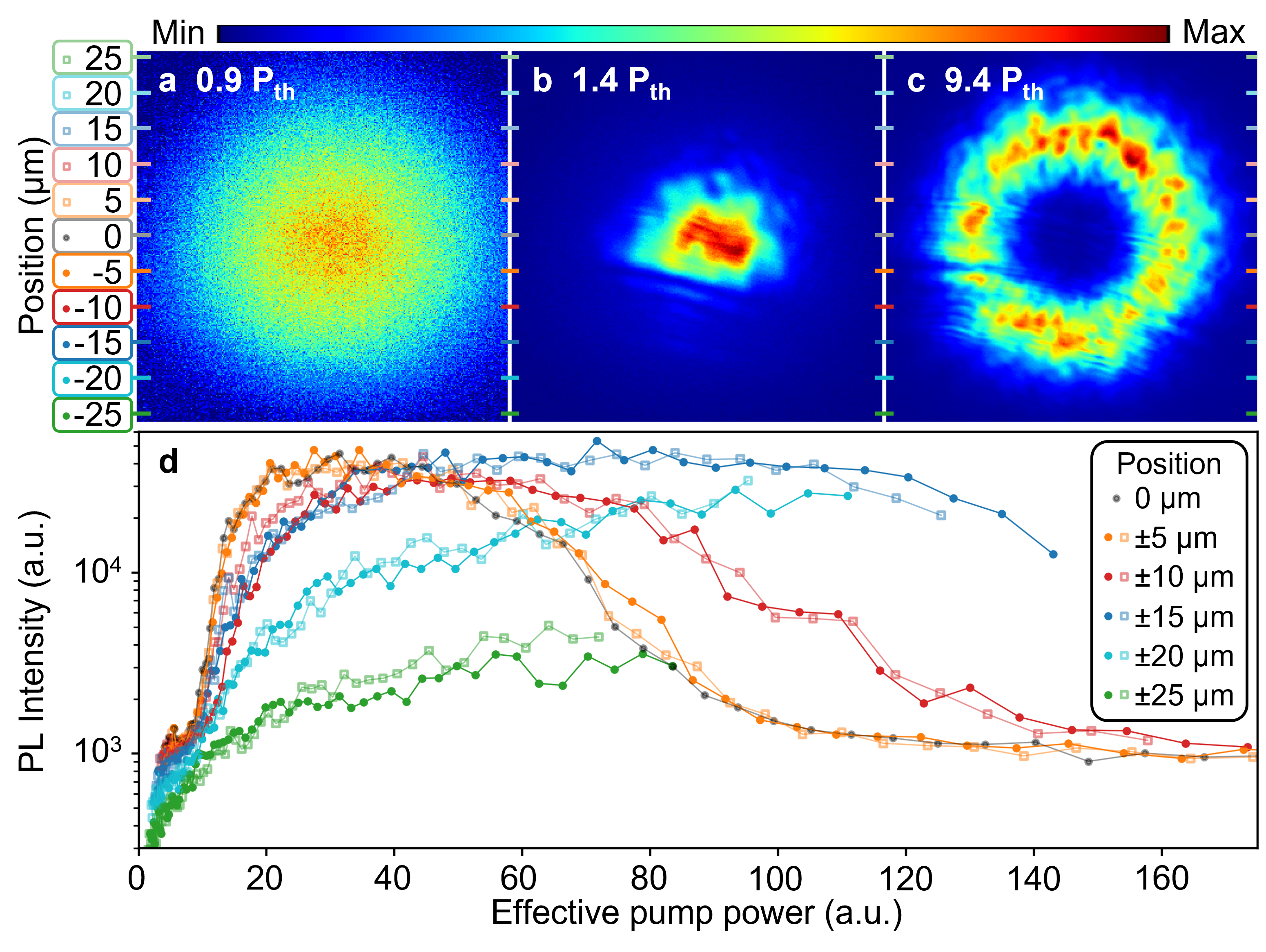}
\vspace{-12pt}
\caption{
Real-space evolution of the polariton condensate.  
(a–c) Emission profiles under Gaussian pumping at different excitation powers: below threshold (a), just above threshold ($\sim$1.4$P_{\mathrm{th}}$, b), and at high power ($\sim$9.4$P_{\mathrm{th}}$, c). The condensate first forms at the pump center before redistributing outward into a reversible annular profile.  
(d) Power-dependent emission intensity extracted at increasing radial distances from the pump center. The central emission saturates at high fluence, while off-center regions brighten with progressively higher thresholds, consistent with repulsive polariton–reservoir and polariton-polariton interactions driving outward flow.  
}
\label{fig:4}
\vspace{-5pt}
\end{figure}

To probe the condensate formation, we imaged the real-space emission profiles (Fig.~\ref{fig:4}). Below threshold, the emission inherits the Gaussian pump profile (Fig.~\ref{fig:4}a). Just above threshold ($\sim$1.4$P_{\mathrm{th}}$), the emission collapses into the center of the spot, forming a condensate, consistent with the higher excitation density in that region (Fig.~\ref{fig:4}b)\cite{Daskalakis2015,Bobrovska2018}. At higher excitation powers, however, the condensate redistributes outward as it gains momentum (0.48 $\mu$m$^{-1}$ in Fig.~\ref{fig:3}e), forming an annular emission profile (Fig.~\ref{fig:4}c). 
We verified that this redistribution is not an artifact of sample degradation. As shown in Supplementary Fig.~\ref{S_fig:ASE_damage}, neat DPAVB:PS films exhibit irreversible damage under ASE cycling, whereas the polariton microcavities can be repeatedly driven across threshold without any change in emission profile or intensity (Supplementary Fig.~\ref{S_fig:polariton_no_damage}). This confirms that the observed spatial  profiles are reversible—contrasting with the irreversible rings observed in bare films—and therefore intrinsic to the condensate rather than arising from material or optical damage. We next quantify this redistribution by analyzing the spatially resolved input–output characteristics.


To quantify this behaviour, we extracted input–output curves at different radial distances from the pump center (Fig.~\ref{fig:4}d). The absorbed pump fluence was renormalized by the Gaussian intensity distribution of the excitation spot below threshold (see Methods), and is hereafter referred to as the \emph{effective pump power}. Close to the center, the emission shows a rapid nonlinear increase up to an effective pump power of about 20 (arbitrary units), similar to the spatially-integrated threshold in Fig.~\ref{fig:3}a. Beyond this, the central emission saturates, while off-center regions continue to brighten, with their effective thresholds shifting progressively outward (e.g. $\sim$60 a.u. at +10~µm, $\sim$80~a.u. at +15~µm). This spatially resolved behaviour, together with the momentum redistribution observed in Fig.~\ref{fig:3}, is consistent with repulsive polariton–polariton and polariton–reservoir interactions driving the condensate outward, adding to the emission intensity at the edges of the pumped region. Comparing the radial intensity profiles of the real-space emission rings of three differently detuned cavities shows how the ring radius becomes larger when the excitonic content of the LP increases (Supplementary Fig.~\ref{S_fig:ring_detunings}).

We interpret this as evidence of a new polariton lasing channel in which the condensate depopulates the pump center and occupies regions of lower exciton density. This outward redistribution not only reflects repulsive polariton interactions but also provides a mechanism to avoid exciton saturation at the pump center, thereby protecting against degradation of strong coupling. We note that related outward-flow and annular condensation phenomena have been reported in inorganic microcavities such as GaAs and ZnO, often aided by engineered traps or ring-shaped pumping geometries. However, to our knowledge, an annular redistribution emerging under Gaussian nonresonant excitation has not previously been observed in organic microcavities~\cite{hahe_interplay_2015, ballarini_self-trapping_2019, kalevich_ring-shaped_2014, Christmann2012}.

\begin{figure}
\vspace{-5pt}
\centering
\includegraphics[width=\linewidth]{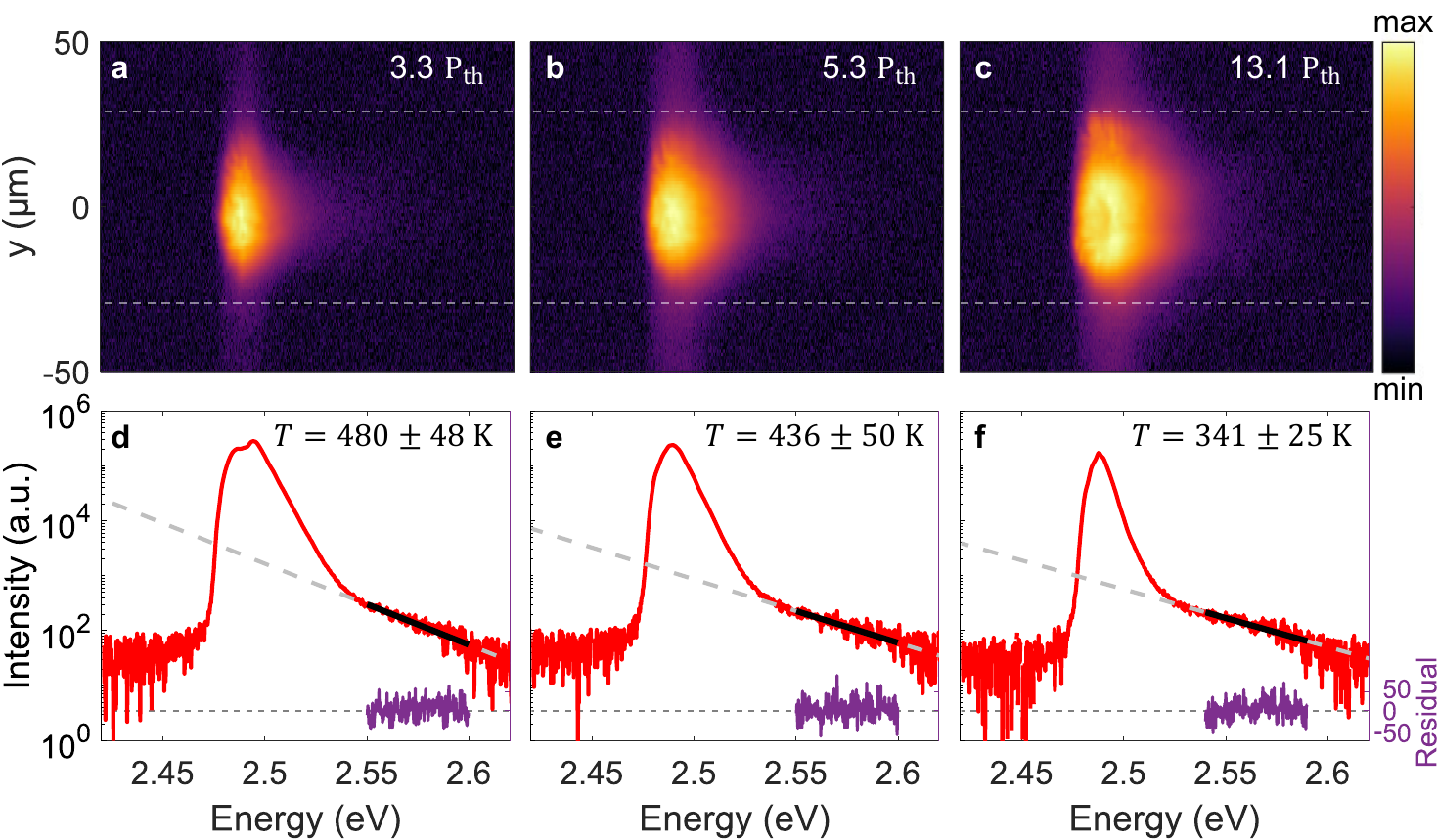}
\vspace{-12pt}
\caption{Thermalization of the polariton condensate.  
(a–c) Real-space resolved emission spectra under Gaussian pumping at different excitation powers above threshold: a) $\sim$3.3$P_{\mathrm{th}}$, b) $\sim$5.3$P_{\mathrm{th}}$, and c) $\sim$13.1$P_{\mathrm{th}}$.   
(d-f) Fits of the thermal tail to the Maxwell-Boltzmann distribution, with the obtained fit temperatures indicated along with 95\% confidence bounds. The solid black lines mark the data range included in the least-squares fitting. The data (red) are obtained by integrating over the spatial axis between the horizontal gray lines in (a-c), from -30µm to 30µm. Fit residuals are plotted in linear scale on the right‑hand y‑axis in the same (arbitrary) intensity units as the spectra. For root‑mean‑squared error (RMSE) and fit window sensitivity analysis, see Supplementary Fig.~\ref{S_fig:fit_analysis}.}
\label{fig:5}
\vspace{-5pt}
\end{figure}

Lastly, we examined the thermalization properties of the condensate. As the excitation power increases from $\sim3.3$ $P_{\mathrm{th}}$ to$\sim13.1$ $P_{\mathrm{th}}$, the emission spectra in Fig.~\ref{fig:5}(a–c) show an increasingly well-defined thermal tail at higher energies (linear slope in the semi-logarithmic scale). Fitting this tail to a Maxwell–Boltzmann distribution (Fig.~\ref{fig:5}(d–f)) reveals a clear decrease in the extracted effective temperature: from 480~K at $\sim3.3$ $P_{\mathrm{th}}$, to 436~K at $\sim5.3$ $P_{\mathrm{th}}$, and down to 341~K at $\sim13.1$ $P_{\mathrm{th}}$. This trend suggests progressive thermalization of the polariton population with increasing excitation density. We fit the Maxwell–Boltzmann distribution exclusively to the high‑energy tail of the photoluminescence spectrum, where the Bose–Einstein distribution reduces to its Maxwell–Boltzmann limit, thereby avoiding the divergence near the ground‑state peak and enabling a robust extraction of the effective temperature. We note that the temperature extracted from the Maxwell–Boltzmann fit should be understood as a phenomenological slope parameter rather than evidence of global thermal equilibrium. The observed cooling of the high-energy tail is likely driven by polariton–polariton and polariton-reservoir interaction and the resulting scattering processes occurring predominantly near the center of the cavity, where the density is highest. Based on previous studies of polariton condensates, in which polariton-polariton interactions and scattering have been identified as crucial mechanisms for establishing thermalization and condensation~\cite{sun_direct_2017,yoon_enhanced_2022,alnatah_coherence_2024,gomez-dominguez_materials_2025}, we speculate that a similar process is responsible here. Further studies will be required to determine the underlying mechanisms in detail. The data in Fig.~\ref{fig:5}(d–f) were obtained by integrating spatially over the central $\pm$30~µm region, as indicated by the gray horizontal dashed lines in Fig.~\ref{fig:5}(a–c). The fits were performed using the data within the energy range marked by black lines in Fig.~\ref{fig:5}(d–f), spanning 50~meV—approximately twice the thermal energy at room temperature—thus providing a robust fitting range. We investigated the sensitivity of the extracted temperature to the choice of fitting window (Supplementary Fig.~\ref{S_fig:fit_analysis}) and conclude that the observed temperature trend is robust. Moreover, as shown in Supplementary Fig.~\ref{S_fig:fit_spatial}, the extracted fit temperature shows only minor variation with spatial position (center vs. annulus), indicating that the globally integrated spectrum provides a reasonable overall representation of the temperature extracted from each spatial region.



\section{Summary}
In summary, we have demonstrated the first fully solution-processed polariton microcavity laser, fabricated entirely by spin coating of both dielectric mirrors and the active DPAVB:PS layer. 
This advance not only greatly simplifies the fabrication of organic polariton devices but also addresses a fundamental challenge of organic semiconductors: their intrinsic instability\cite{Diesing2024}. 
While improving the chemical robustness of organics remains a long-standing challenge—as exemplified by OLED research—a pragmatic workaround is to enable device architectures that can be fabricated rapidly, with low energy cost, and at scale\cite{zhao2024stable,mischok2024breaking}. 
Our approach follows this strategy by delivering high-quality mirrors and cavities using solution processing alone. Rather than positioning solution processing as intrinsically superior to vapour-based fabrication, the aim here is to establish it as a viable and experimentally accessible alternative for realizing high-quality organic polaritonic microcavities.

Beyond this technical advance, our platform has revealed a new polariton condensation pathway: under Gaussian excitation the condensate redistributes outward, forming a reversible annular profile not previously observed in organic systems. Additionally, we found that with increasing excitation power, the high-energy tail of the emission spectra becomes increasingly thermalized, as evidenced by a systematic decrease in the extracted effective temperature from 480~K to 341~K. This cooling trend indicates enhanced polariton and exciton scattering at high densities.

Although a comprehensive study of this novel physical mechanism is planned, we attribute this to repulsive polariton–polariton and polariton-reservoir interactions that depopulate the pumped center and feed the periphery, thereby mitigating exciton saturation and protecting strong coupling at high excitation densities. 
However, several alternative scenarios may also be relevant in the future~\cite{Lerario2017,Xu2023,Pandya2021,Balasubrahmaniyam2023,siltanen2024incoherentpolaritondynamicsnonlinearities}, further contributing to the ongoing debate on polariton filtering~
\cite{Cheng2022,George2024,Schwennicke2025}.
The observed interaction-driven redistribution and enhanced scattering highlight the role of polaritons in reshaping relaxation pathways and polariton filtering in organic semiconductors~\cite{Lerario2017,Xu2023,Pandya2021,Balasubrahmaniyam2023,siltanen2024incoherentpolaritondynamicsnonlinearities,Cheng2022,George2024,Schwennicke2025}. In particular, the ability of polaritonic states to modify excited-state dynamics is central to ongoing efforts to mitigate the accumulation of long-lived triplet excitons, which currently limits electrically injected organic lasing. By enabling fully solution-processed DBR microcavities that support polariton lasing, the present platform lowers the experimental barrier for exploring polariton-assisted strategies to address triplet-related limitations.

Together, these results highlight how simple, energy-efficient fabrication, combined with tailored material choice, can unlock qualitatively new polariton physics and provide a foundation for scalable classical and quantum photonic applications, and may help accelerate progress toward electrically driven organic lasers\cite{Giebink2008,Yoshida2023}.

\section{Methods}

\subsection{Materials and Thin film fabrication}
High-refractive-index titanium hydroxide/polyvinyl alcohol (TiOH/PVA) hybrid films were synthesized using titanium(IV) chloride (TiCl\textsubscript{4}, quality level 100) and polyvinyl alcohol (PVA, MW 30,000–70,000), both obtained from Sigma-Aldrich. The TiOH stock solution was prepared by slowly hydrolyzing 2.2 mL of TiCl\textsubscript{4} in 20 mL of cold deionized water under continuous stirring in an ice bath. This TiOH solution was subsequently mixed with an aqueous PVA solution in a 1:1 volume ratio to form the TiOH/PVA hybrid precursor, following the procedure reported by \cite{Russo2012}. The hybrid films were then deposited using a Laurell spin coater via a single-step spin coating and annealing process.
Low-refractive-index layers were fabricated from Nafion by diluting a commercial Nafion D520 dispersion (5\% in water and low aliphatic alcohols, Chemours) to the desired concentration in isopropyl alcohol (IPA, Sigma-Aldrich). The resulting solutions were spin-coated using the same setup to obtain uniform Nafion films.
The active DPAVB films were prepared by dissolving polystyrene (PS) into toluene at a concentration of 30 mg/mL, followed by the addition of 20 mg of DPAVB (Luminescence Technology) to obtain the DPAVB:PS solution. The mixed solution was spin-coated and annealed to form the DPAVB active films.

\subsection{Sample fabrication}
The substrates (15 mm × 15 mm × 1 mm) were cleaned sequentially with a 3\% Decon 90 water/soap solution and isopropanol to remove any surface residues. Each cleaning step involved 10 minutes of sonication, followed by drying with a nitrogen purge. The same cleaning protocol was applied to the silicon substrates used for determining optical constants and film thicknesses by ellipsometry \cite{Leppala2024}.
As illustrated in Fig.~\ref{fig:1}a, microcavities composed of two DBRs separated by an active layer were fabricated on quartz substrates. During each deposition step, the substrate was spin-coated with the desired solution at 5000 rpm, followed by drying on a hot plate at 80 °C for 30 seconds. Before the subsequent coating, the sample was allowed to cool for one minute at ambient temperature to dissipate residual heat. This process was repeated until the required number of layers was achieved.  The active layer (260 nm thick) was then spin-coated from the DPAVB:PS solution at 2000 rpm and annealed at 80 °C for two minutes. All depositions were performed at 21 °C and 45\% relative humidity.
Unlike our previous work\cite{Qureshi2025}, where additional protective layers and encapsulation steps were required to prevent damage to the organic active layer due to polar solvents, in the present study, no such protection was necessary. This is because the polystyrene host matrix is resistant to the polar solvents used in the DBR fabrication process, allowing direct deposition of the top DBR without compromising the integrity of the active layer.

\subsection{Optical Characterization}
The optical constants and film thicknesses were determined using a J.A. Woollam VASE ellipsometer equipped with a xenon lamp covering a spectral range of 250–2500 nm. The acquired data were analyzed by fitting a Cauchy model in the transparent region of the films. The dispersion characteristics of the DBRs and polaritonic modes were measured using the same ellipsometer setup, where collimated light from a halogen lamp illuminated the sample through a 0.75 NA microscope objective.

PL measurements were performed by exciting the sample at normal incidence from the substrate side using 250 fs pulses at 380 nm with a 0.2 kHz repetition rate (Light Conversion Pharos, Orpheus, and Lyra). The emitted PL was collected through a Nikon 0.75 NA microscope objective and directed into a Teledyne HRS-300 spectrometer equipped with a 1200 g/mm grating and Pixis CCD detector (1340 x 400 pixels), enabling angular-resolved detection from approximately $-48^\circ$ to $+48^\circ$ with 0.117 nm CCD resolution. The entrance slit of the spectrometer was set to  10 µm to define the spectral and angular resolution. Time-resolved PL measurements were performed using the same spectrometer coupled to a PicoQuant TimeHarp 260 time-correlated single-photon counting (TCSPC) module, providing a temporal resolution of 200 ps.


Interference measurements were performed using a Michelson interferometer in retroreflector configuration, as described in detail in our previous work~\cite{Daskalakis2015,Moilanen2021}.


The spatial power dependence was studied to investigate the origin of the lasing hole emerging at high pump powers. First, the spatial distribution of the pump fluence was determined by fitting a Gaussian distribution to the real-space emission below the lasing threshold, as in Fig.~\ref{fig:4}a. The effective pump powers for each position were then determined by scaling the total pump power by the relative intensity of the fitted Gaussian at the given position. Each of the power dependence curves in Fig.~\ref{fig:4} d) was formed from 70 real-space images of sample emission, corresponding to different input pump powers. 
Real-space emission images for studying thermalization properties were acquired using the same setup employed for k-space spectroscopy, with a minor modification. An additional imaging lens was inserted before the entrance slit of the spectrometer to project the real-space plane of the sample onto the CCD detector. The spectrometer slit was fully opened to collect the entire emission area, thereby recording the spatial intensity distribution without spectral dispersion.

\subsection*{Author Contributions}
KSD conceived the project. KSD and HAQ designed the structures and guided the experiments. HAQ and AK performed all the fabrications and HAQ and HL all characterization of the samples. AJM performed the thermalization analysis. HAQ, HL, AJM and KSD wrote the manuscript. All authors contributed to the discussion and analysis of the data.

\subsection*{Conflicts of interest}
The authors declare no conflicts of interest.
\newline


\begin{acknowledgement}
This project has received funding from the Research Council of Finland project “X-SHIELD” (Decision number 369819), the European Research Council (ERC) under the European Union’s Horizon 2020 research and innovation programme (grant agreement No. [948260]). Views and opinions expressed are, however, those of the authors only and do not necessarily reflect those of the European Union. Neither the European Union nor the granting authority can be held responsible for them. The authors also acknowledge Materials Research Infrastructure (MARI) at the Department of Physics and Astronomy, University of Turku, for access and support with the broad-ion beam and SEM facilities, and thank Ermei Mäkilä for providing SEM measurements. HAQ received funding from the Finnish Cultural Foundation and the Magnus Ehrnrooth Foundation, and HL from the Ulla Tuominen Foundation. AJM acknowledges support from the Research Council of Finland (Fellowship decision number 368244). We thank Santeri Kanerva for assistance with figure design. 
\end{acknowledgement}

\bibliography{references}
\newpage

\newpage
\setcounter{equation}{0}
\setcounter{figure}{0}
\setcounter{table}{0}
\setcounter{page}{1}
\makeatletter
\renewcommand{\theequation}{S\arabic{equation}}
\renewcommand{\thefigure}{S\arabic{figure}}

\begin{center}
\textbf{\Large Supplementary Information}\\
\end{center}
\begin{center}
\textbf{\Large{A fully solution-processed organic microcavity laser in the strong light–matter coupling regime}}\\
\end{center}
\noindent
Hassan A. Qureshi, Henri Lyyra, Akseli Korkeamäki, Oskar Tuomi, Antti J. Moilanen, and Konstantinos S. Daskalakis\\
\noindent
Corresponding author: konstantinos.daskalakis@utu.fi
\section*{Contents}

\begin{figure}
\vspace{-5pt}
\centering
\includegraphics[width=0.5\linewidth]{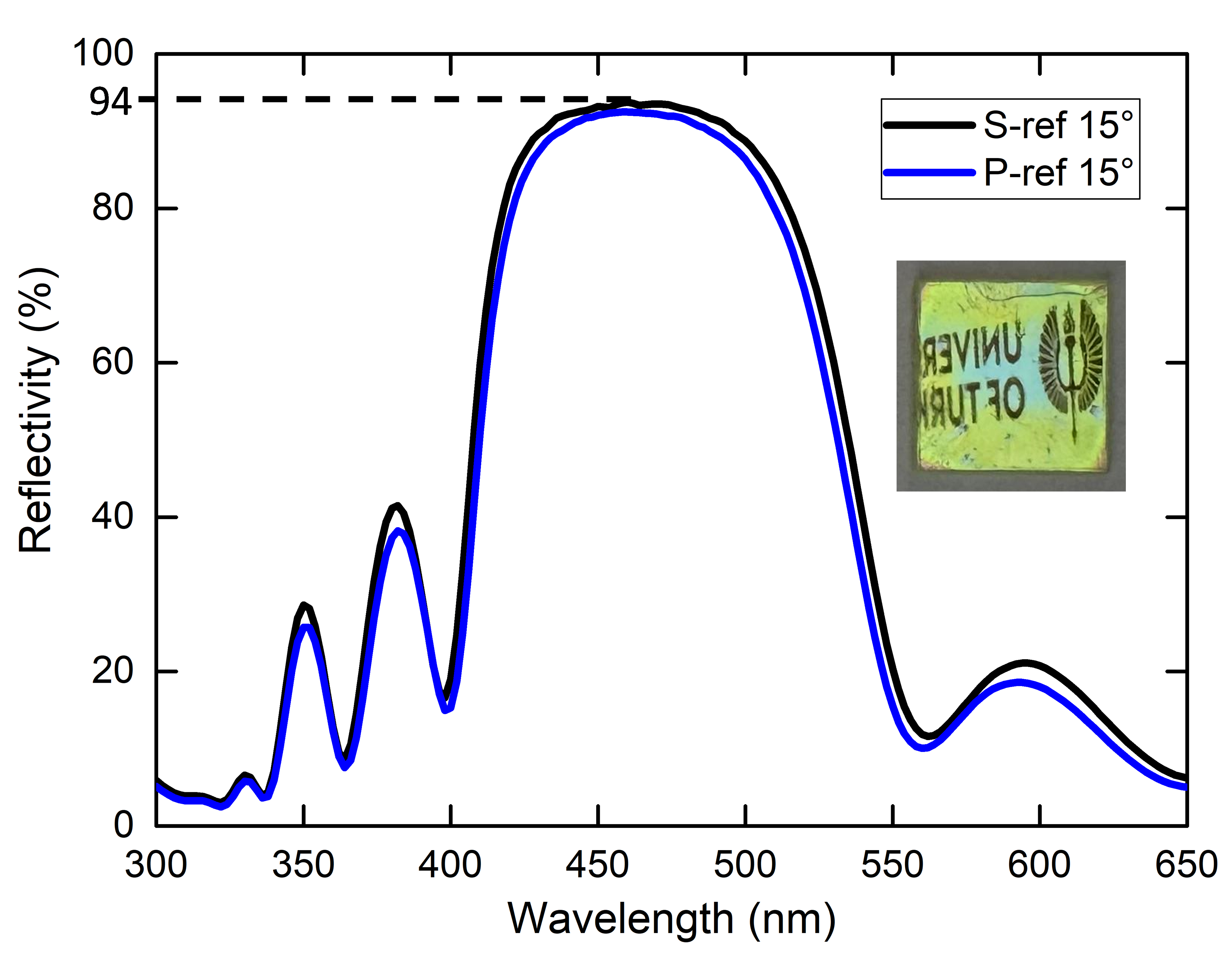}
\vspace{-12pt}
\caption{Reflectivity spectra of the a DBR measured at 15° incidence for s- (black) and p-polarized (blue) light. Both polarizations exhibit similar spectral features with high reflectivity in the stopband region around 480 nm. 
The inset shows the image of as-prepared 7.5 pairs DBR.}
\label{S_fig:DBR_ref}
\vspace{-5pt}
\end{figure}
\begin{figure}
\vspace{-5pt}
\centering
\includegraphics[width=0.5\linewidth]{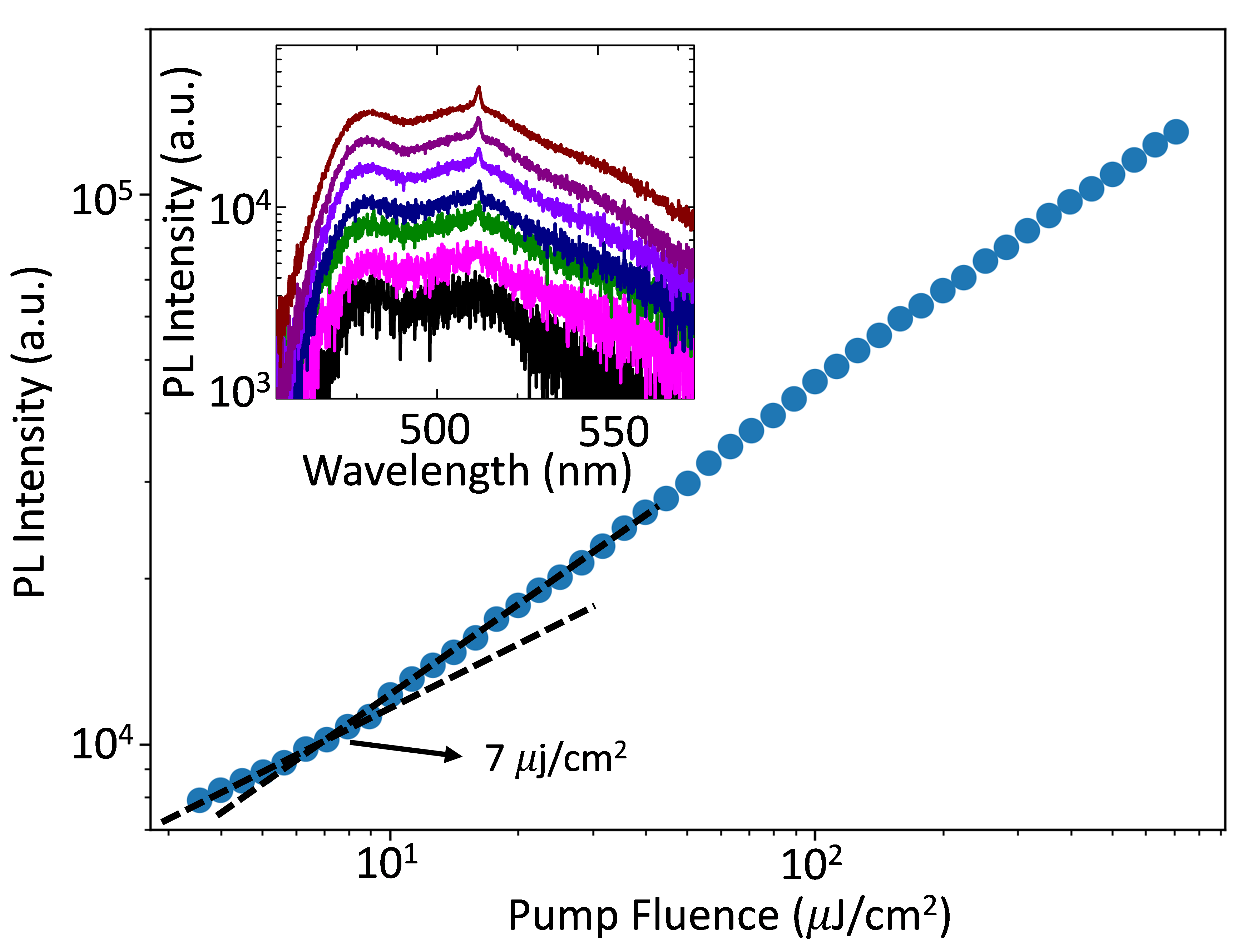}
\vspace{-12pt}
\caption{Power-dependent photoluminescence (PL) measurements of the bare film. The main plot shows the integrated PL intensity as a function of pump fluence, exhibiting a nearly linear increase without a clear threshold behavior. The inset presents the corresponding PL spectra at different fluences, showing amplified spontaneous emission (ASE) with increasing pump power.}
\label{S_fig:ASE_PD}
\vspace{-5pt}
\end{figure}
\begin{figure}
\vspace{-5pt}
\centering
\includegraphics[width=\linewidth]{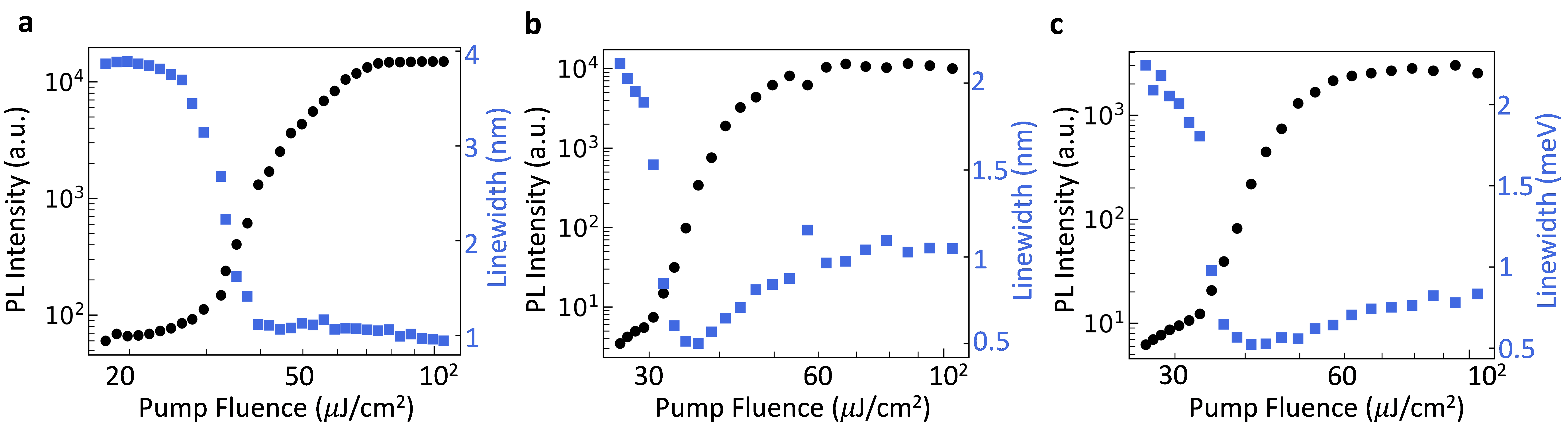}
\vspace{-12pt}
\caption{Power-dependent measurements showing the integrated PL intensity (black dots) and emission linewidth (blue squares) as a function of pump fluence for different pump–collection polarization configurations: (a) s-polarized pump and p-polarized collection, (b) p-polarized pump and p-polarized collection, and (c) p-polarized pump and s-polarized collection. All configurations exhibit a clear threshold-like increase in PL intensity accompanied by linewidth narrowing, characteristic of a polariton lasing. No significant differences in threshold and linewidth are observed between polarization configurations, indicating negligible cavity anisotropy or birefringence. 
}
\label{S_fig:more_lasing_info}
\vspace{-5pt}
\end{figure}
\begin{figure}
\vspace{-5pt}
\centering
\includegraphics[width=0.5\linewidth]{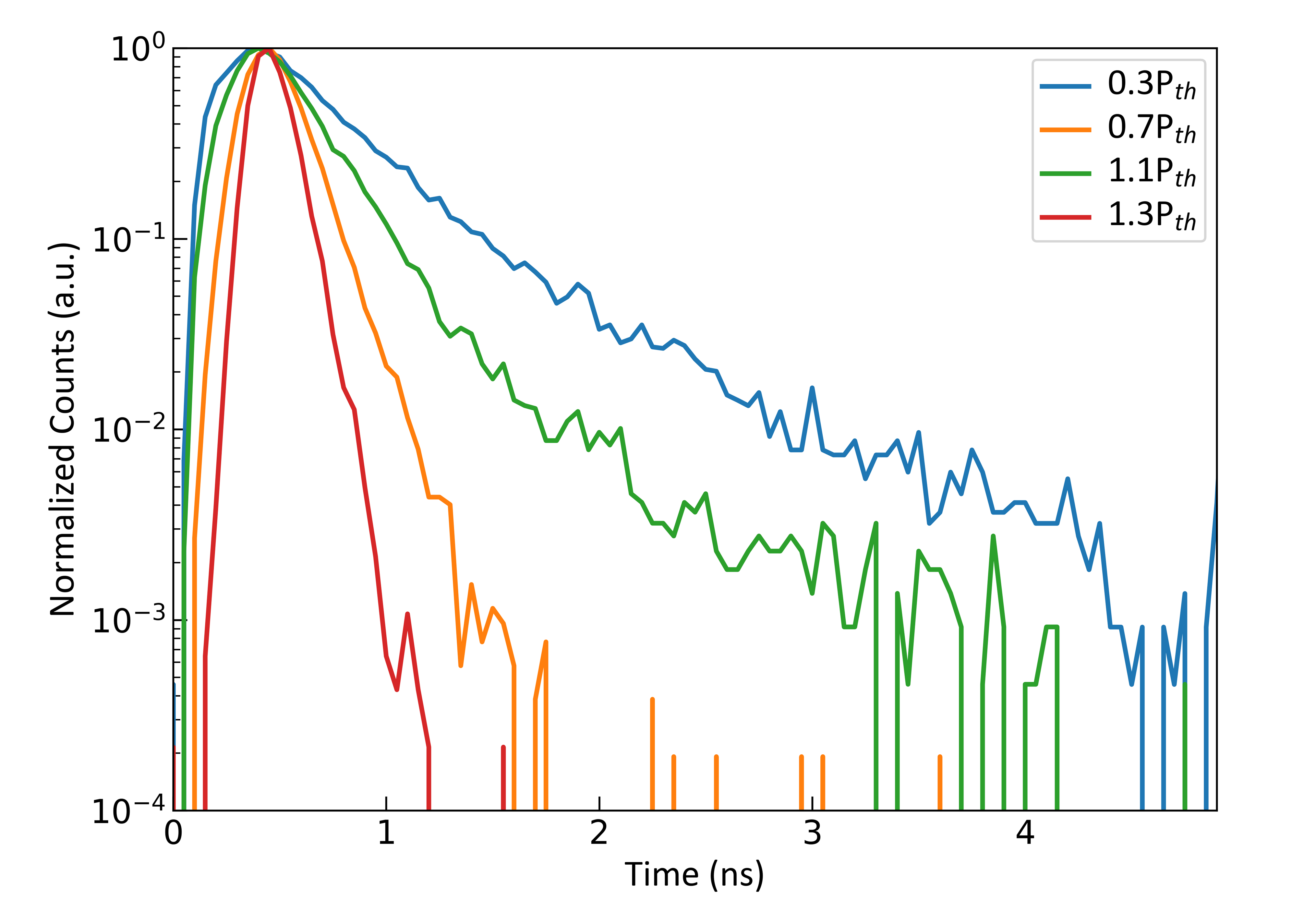}
\vspace{-12pt}
\caption{Time-resolved photoluminescence (TRPL) decay curves of the microcavity emission recorded at different excitation powers. As the pump fluence increases from 0.3 Pth to 1.3 Pth, the decay time shortens significantly indicating accelerated relaxation dynamics as the system transitions to the lasing regime.}
\label{S_fig:transient}
\vspace{-5pt}
\end{figure}
\begin{figure}
\vspace{-5pt}
\centering
\includegraphics[width=\linewidth]{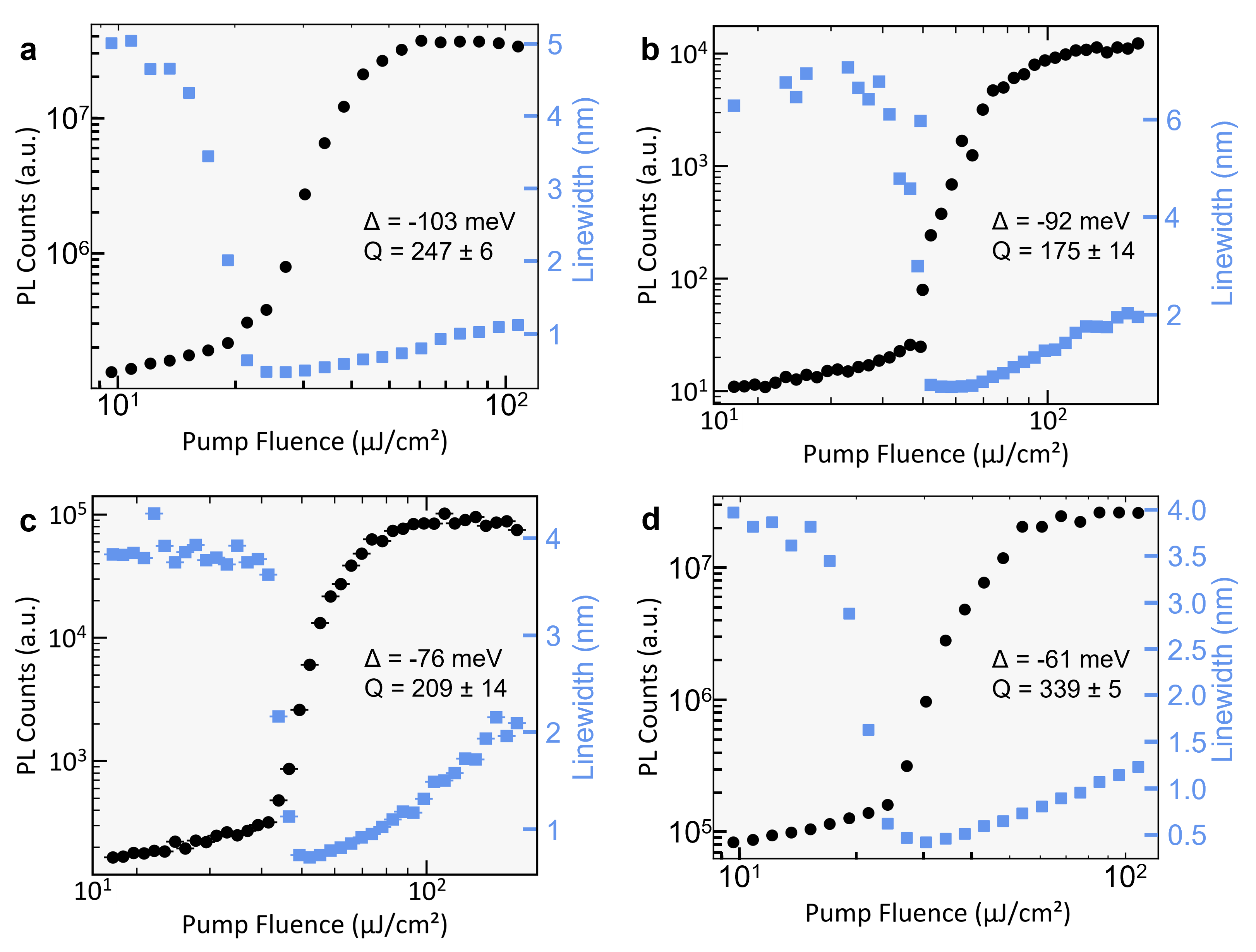}
\vspace{-12pt}
\caption{
Power-dependent PL characteristics with emission peaks at (a) 511.25 nm, (b) 510 nm, (c) 508 nm, and (d) 506.25 nm. 
Panels (a - d) show the integrated PL intensity (black) and linewidth (blue) as functions of pump fluence, exhibiting a clear threshold-like behavior and linewidth narrowing, indicative of the transition to the polariton lasing regime. $\Delta$ is the exciton-cavity detuning and Q is determined from five low-fluence measurements integrated over small collection angles. The data in panels (a) and (d) are from the same measurements as the data in Figs.~\ref{S_fig:dispersion_detunings} and \ref{S_fig:ring_detunings} of the cavity detunings $\Delta =$ -103 meV and $\Delta =$ -61 meV, respectively.
}
\label{S_fig:Lasing_in_other_det}
\vspace{-5pt}
\end{figure}
\begin{figure}
\vspace{-5pt}
\centering
\includegraphics[width=0.5\linewidth]{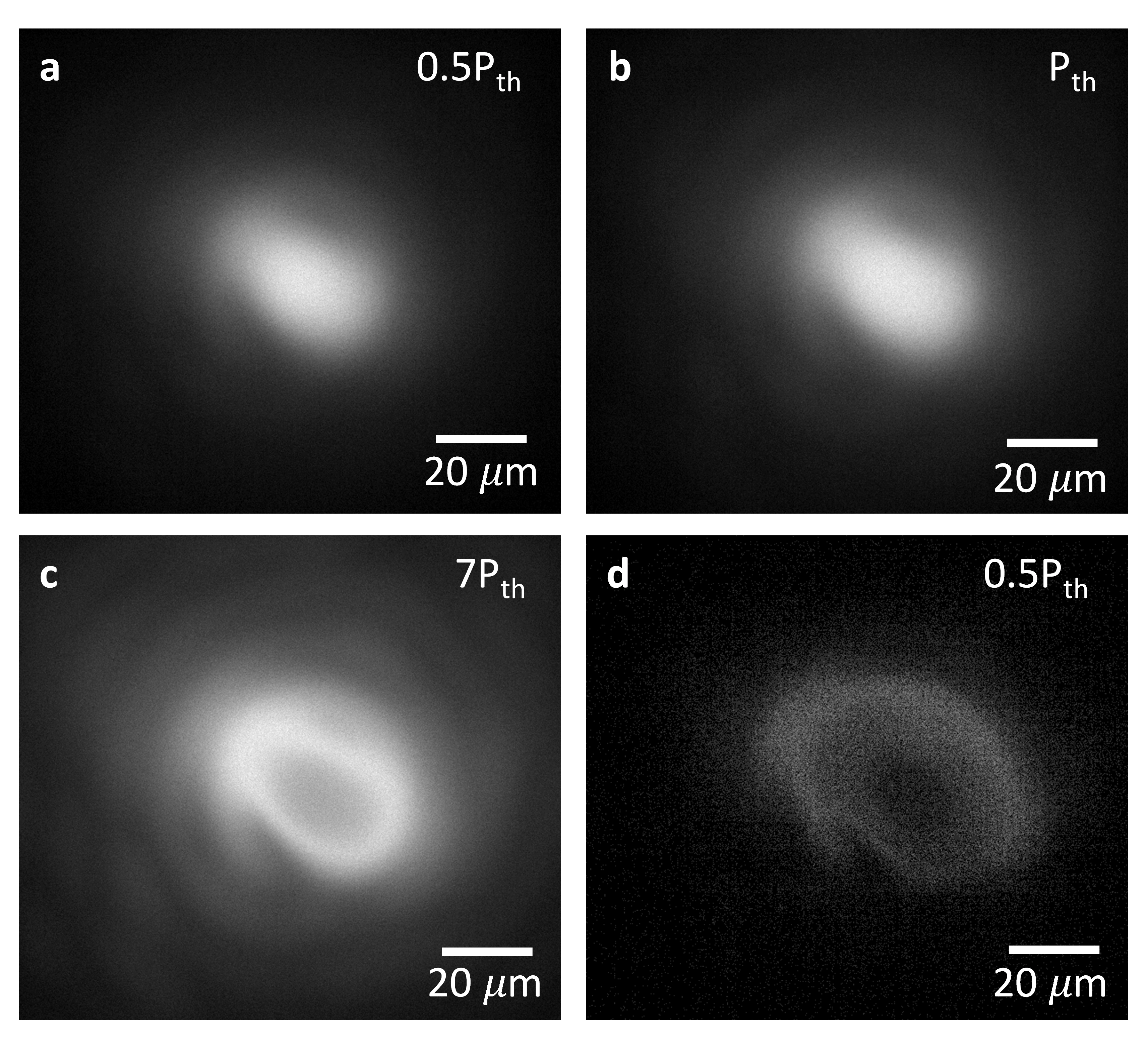}
\vspace{-12pt}
\caption{Real-space emission images of the film recorded at 0.5 P$_{th}$, P$_{th}$, and 7 P$_{th}$, respectively, show the progressive formation of a ring-like feature as the pump power increases. (d) Image of the same spot taken again at 0.5 P$_{th}$ after exposure to higher power (7 P$_{th}$) clearly shows the same ring pattern, confirming that the observed ring in the film originates from irreversible damage compared to the reversible ring shape emission observed in polariton microcavities.
}
\label{S_fig:ASE_damage}
\vspace{-5pt}
\end{figure}
\begin{figure}
\vspace{-5pt}
\centering
\includegraphics[width=0.9\linewidth]{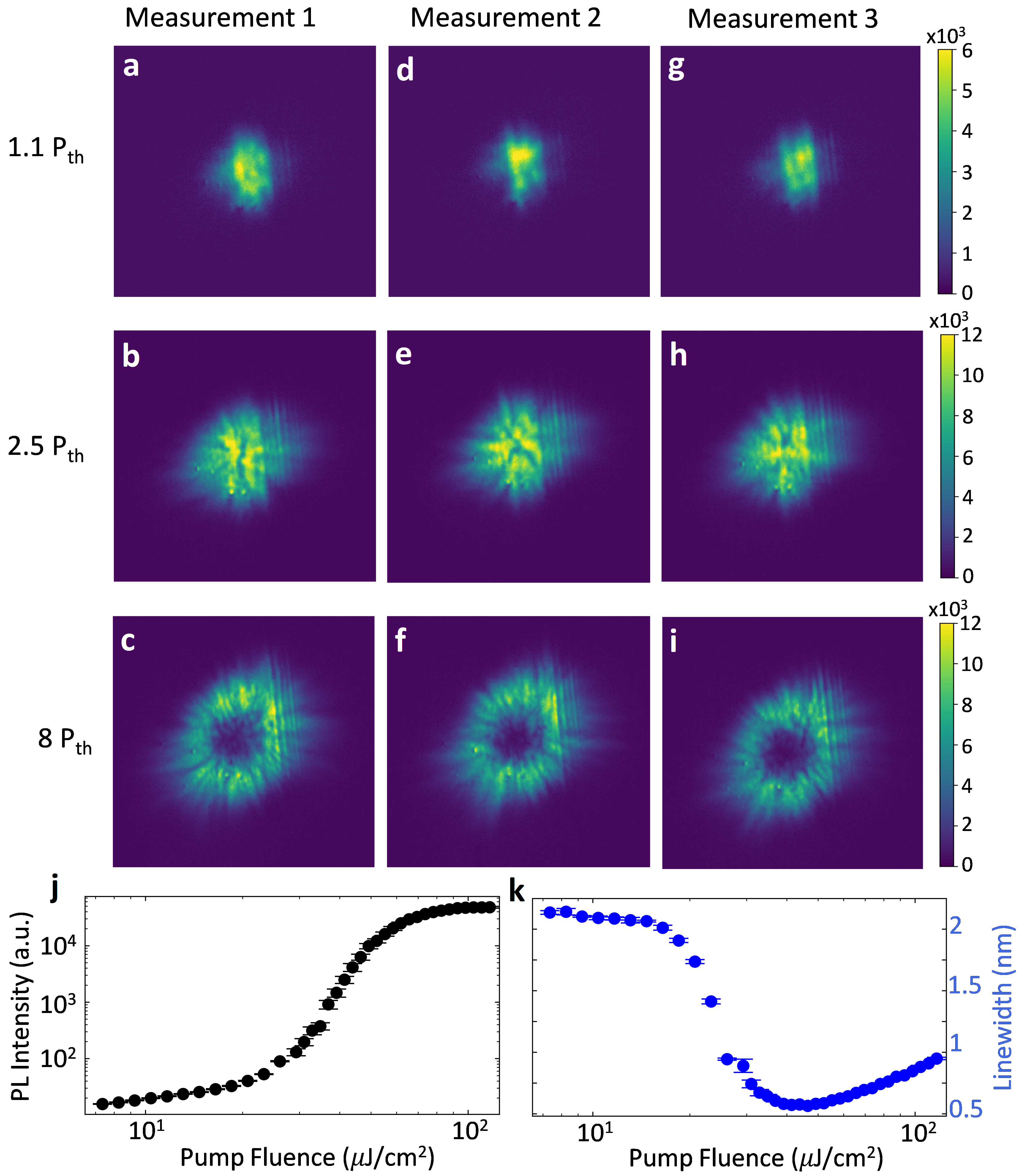}
\vspace{-12pt}
\caption{(a-i)Real-space emission images recorded at the same spot under different excitation powers and repeated three times to verify reproducibility. The measurement order is (a)$\rightarrow$(b)$\rightarrow$(c)$\rightarrow$(d)$\rightarrow$(e)$\rightarrow$(f)$\rightarrow$(g)$\rightarrow$(h)$\rightarrow$(i). At low excitation (1.1 P$_{th}$), localized emission is observed, while at higher powers (8 P$_{th}$), a ring-shaped emission pattern consistently appears in all measurements. The repeatable emergence of the ring structure confirms that it is an intrinsic feature of the condensate rather than damage to the film.
(j,k) Power-dependent LP photoluminescence intensity (black dots) and  corresponding LP linewidth (blue dots). Error bars represent the standard deviation obtained from three independent measurements. The measurements were performed in the same sample spot back to back and the power curves were also done back to back after full power cycles.}
\label{S_fig:polariton_no_damage}
\vspace{-5pt}
\end{figure}
\begin{figure}
\vspace{-5pt}
\centering
\includegraphics[width=0.8\linewidth]{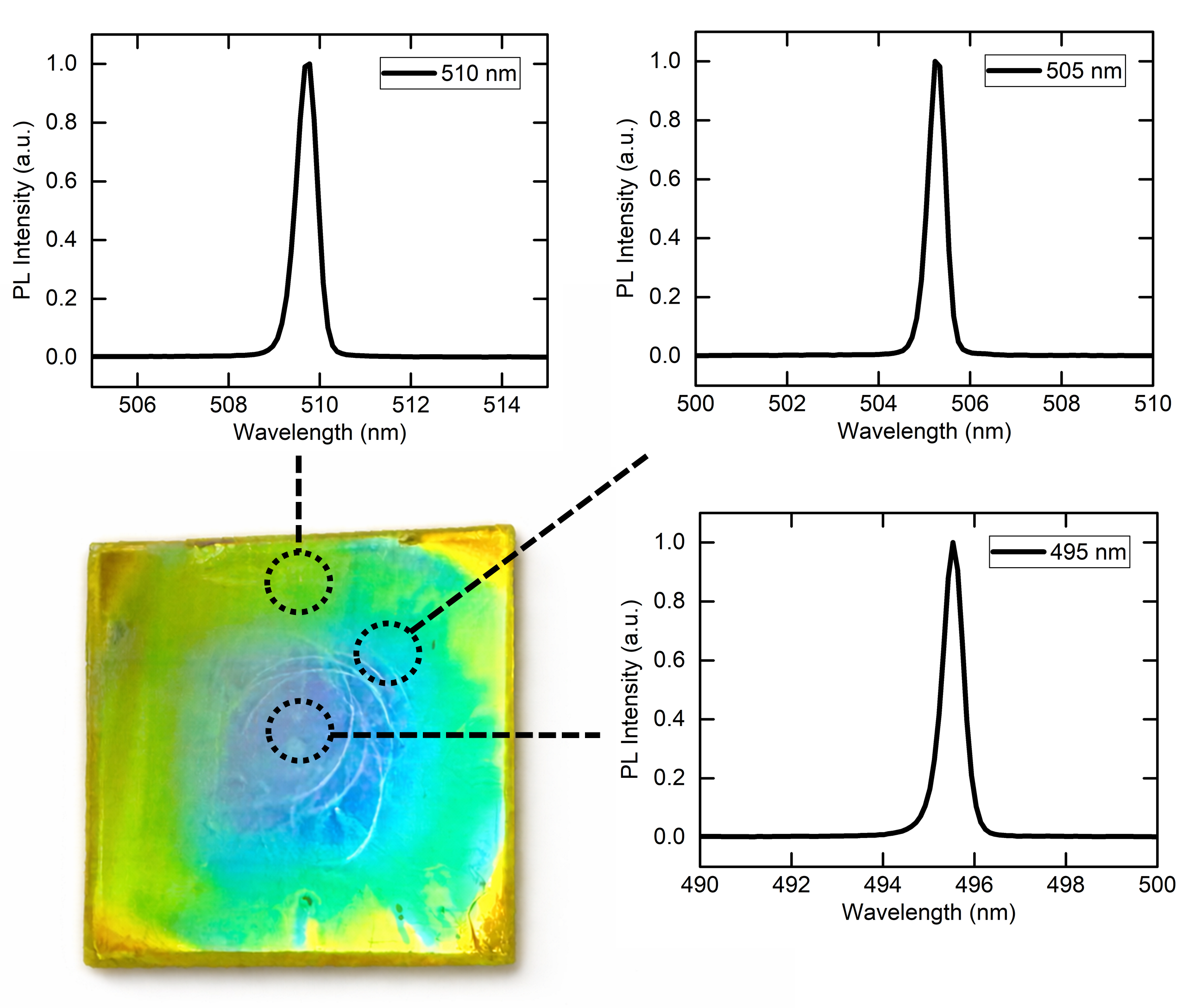}
\vspace{-12pt}
\caption{Photograph of the fabricated microcavity sample illustrating the spatial thickness gradient induced by spin coating. The inherent coating dynamics produce a gradual cavity thickness variation across the substrate, resulting in position-dependent detuning. By translating the collection area across the sample (dashed lines), different detuning conditions can be accessed on a single device. Representative PL spectra recorded at three distinct spatial positions (dotted circles) show clear spectral shifts (495 nm, 505 nm, and 510 nm), directly demonstrating the spatial variation of the cavity resonance.}
\label{S_fig:polariton_photo}
\vspace{-5pt}
\end{figure}
\begin{figure}
\vspace{-5pt}
\centering
\includegraphics[width=\linewidth]{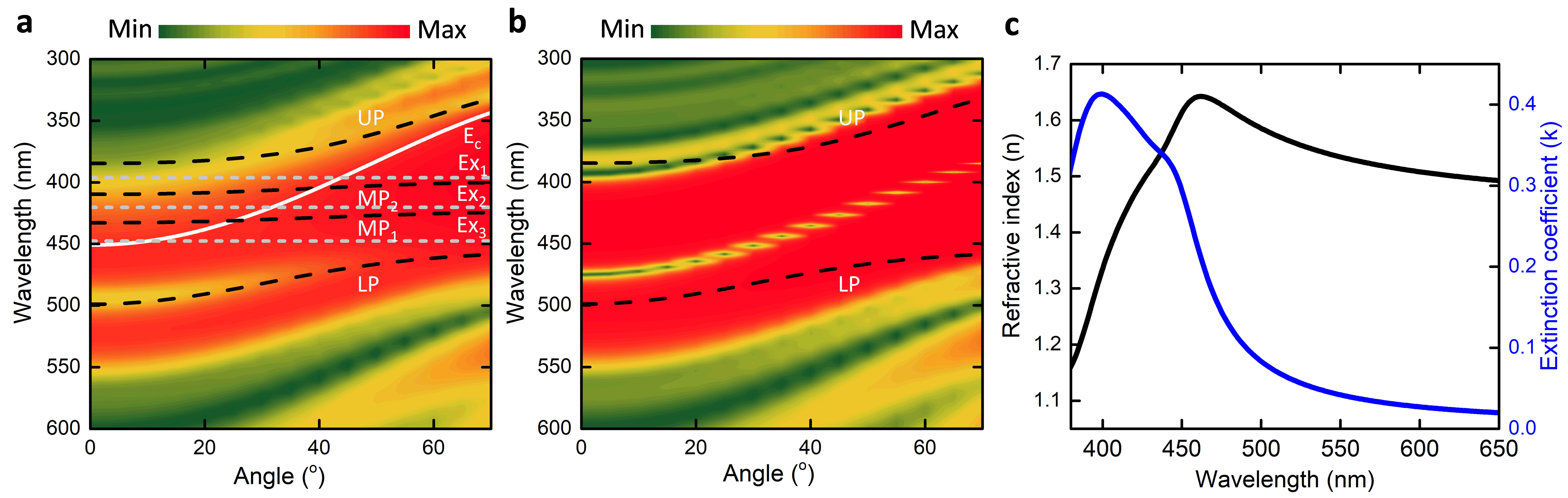}
\vspace{-12pt}
\caption{Transfer-matrix modelling (TMM) of the cavity reflectivity using the experimentally measure refractive index (n and k).
(a) Full-cavity TMM including excitonic resonances, where the coupling parameters are taken from the CHOM fitted to the experimental reflectivity data. The simulation reproduces the observed anticrossing and dispersion curvature, consistent with sustained strong coupling. Dashed lines indicate the calculated polariton branches obtained from the coupled oscillator model (CHO).
(b) TMM of the empty cavity without excitonic resonances, showing only the bare cavity and higher-order Bragg modes. In this case, no anticrossing or mode splitting is observed. Dashed lines mark the dispersion of the bare cavity photon mode for reference.
(c) Experimentally measured complex refractive index (n and k) of the DPAVB:PS film.}
\label{S_fig:Transfer matrix}
\vspace{-5pt}
\end{figure}
\begin{figure}
\vspace{-5pt}
\centering
\includegraphics[width=\linewidth]{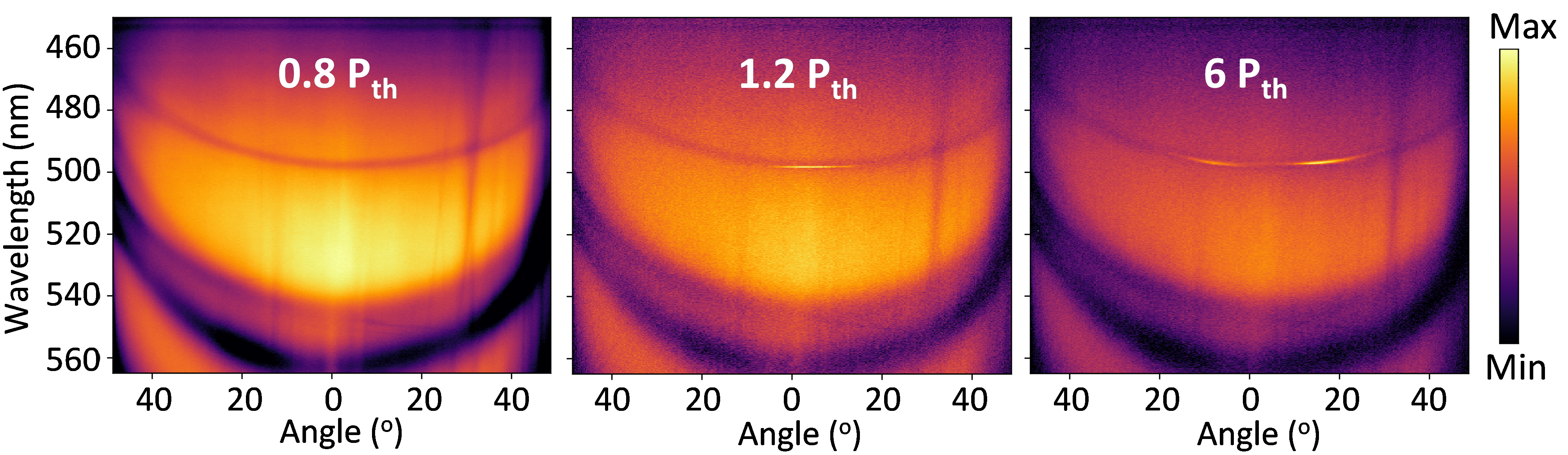}
\vspace{-12pt}
\caption{Steady-state angle-resolved reflection measured below and above threshold at $0.8\,P_{\mathrm{th}}$, $1.2\,P_{\mathrm{th}}$, and $6\,P_{\mathrm{th}}$. The LP dispersion remains spectrally and angularly unchanged across threshold, with no observable distortion or shift of the LP or higher-order Bragg modes. Above threshold, the lasing emission remains strongly confined within the established LP branch and does not evolve toward the bare cavity photon dispersion.}
\label{S_fig:lasing_reflectivity}
\vspace{-5pt}
\end{figure}
\begin{figure*}[h!]
\centering
\includegraphics[width=1.0\linewidth]{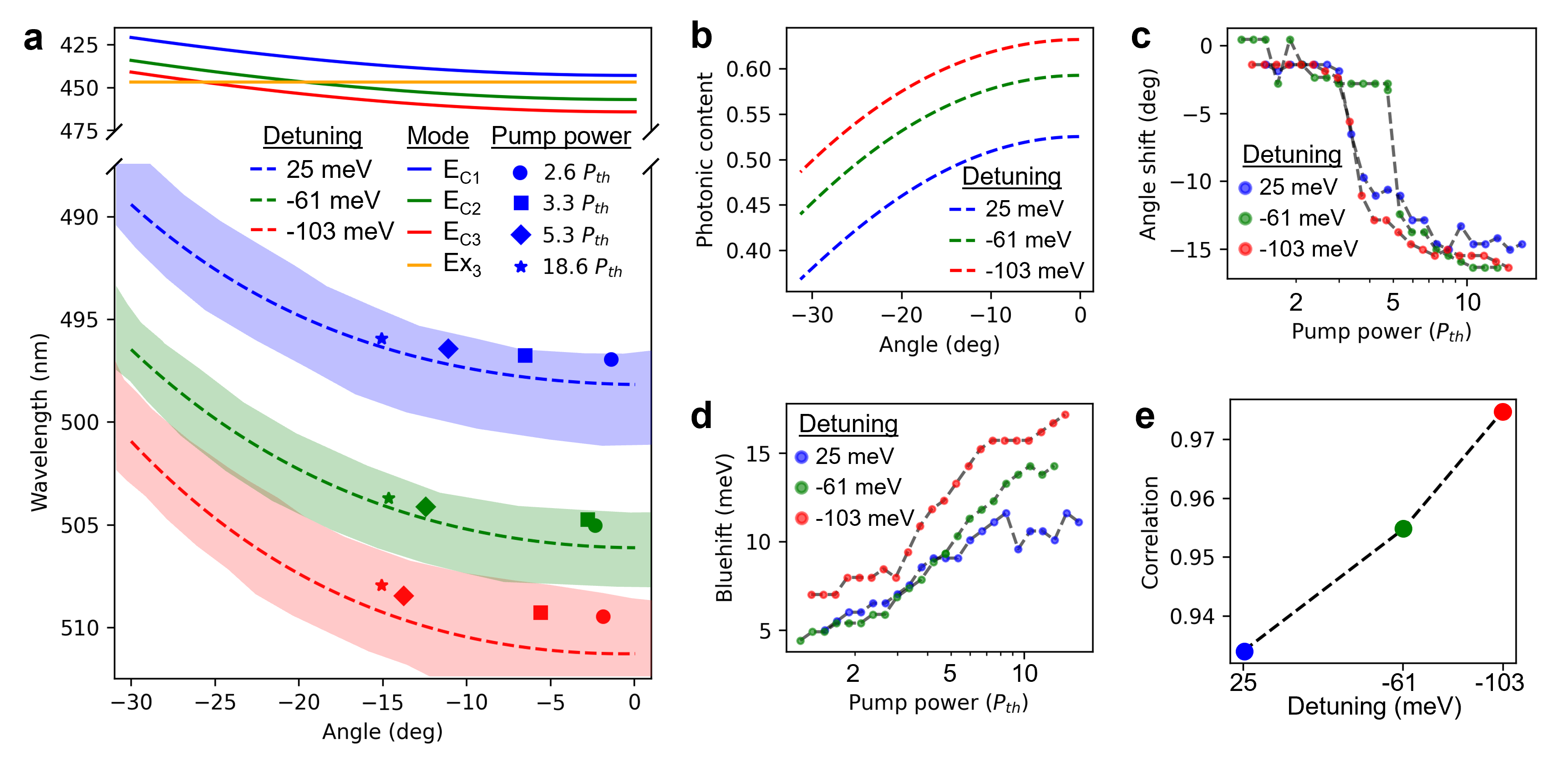}
\caption{The blueshift and angle shift along the below-threshold LP dispersion for three differently detuned cavities. 
(a) Above-threshold emission obeys the below-threshold dispersion. The solid lines are the exciton energy $E_{X_3}$ and the dispersions of the bare cavities ($E_{C1}$, $E_{C2}$, and $E_{C3}$). The dashed lines show the below-threshold dispersions of the LP branches corresponding to the matching-colored cavity dispersions. The shaded areas highlight the parameter regions where the below-threshold EL intensity is at least half of its maximum. The different scatter markers show the highest emission intensity angle-wavelength coordinates at different pump powers, where $P_{th}$ is the lasing threshold. While the pump power increases, the emission undergoes blueshift, but simultaneously the emission angle shifts keeping the highest intensity clearly within the below-threshold dispersion’s linewidth.
(b) The coupled harmonic oscillator model simulation of the Hopfield coefficients shows that increasing the detuning makes the polariton more photonic.
(c) The angle shift and (d) the blueshift of the highest emission coordinate as function of the pump power. (e) The Spearman $\rho$ correlation coefficient between the blueshift and angle shift for the three differently detuned cavities from (c-d). The correlation is normalized between -1 and +1, where -1 means perfect anticorrelation, 0 means uncorrelated properties, and +1 means perfect correlation. Here the correlation increases monotonically with detuning, suggesting stronger dependence between blueshift and dispersion for the more photonic polaritons. The integrated PL intensity and linewidth for the detunings 25 meV, -61 meV, and -103 meV, are shown in Figs.~\ref{fig:3} (a), and \ref{S_fig:Lasing_in_other_det} (c) and (d), respectively.
}
\label{S_fig:dispersion_detunings}
\end{figure*}
\begin{figure*}[h!]
\centering
\includegraphics[width=1.0\linewidth]{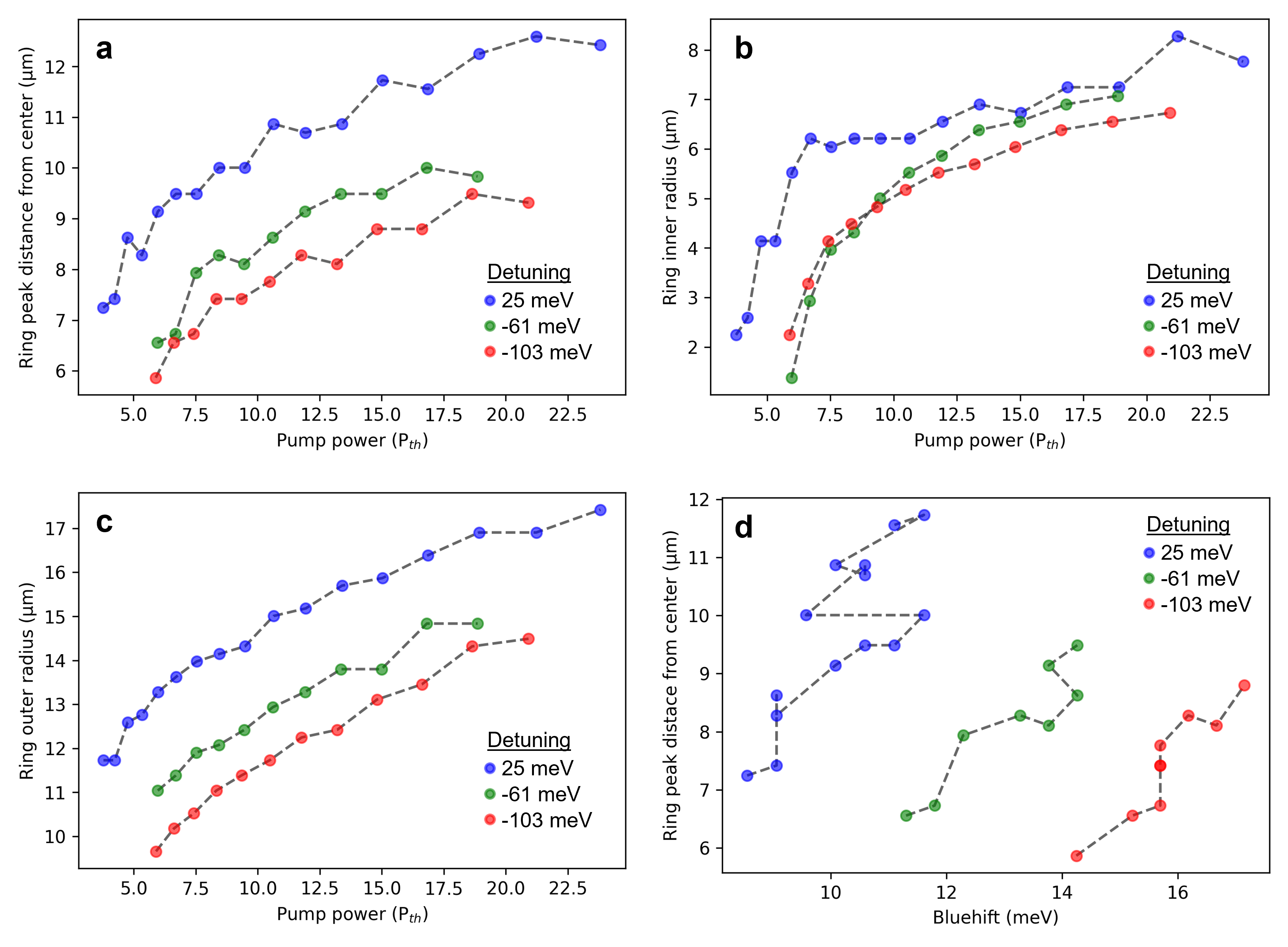}
\caption{The real-space emission ring. As the pump power increases beyond the lasing threshold, a ring-like feature emerges around the pump spot's center in the real-space emission. Further increasing the pump power enlarges the ring as seen in Fig.~\ref{fig:4} of the main article. For each detuning and pump power, the 2D real-space intensity distribution is averaged over all the angles around the pump center position and the resulting radial intensity distribution is analyzed in panels (a-c).  
(a) Ring’s maximum intensity shift, (b) inner half-maximum shift, and (c) the outer half-maximum shift as function of the pump power. The maximum intensity and outer radius shifts are larger for the less-detuned  cases which correspond to higher excitonic polariton content. This suggests that the hole forms due to the repulsive polariton-polariton interaction, which increases with the polariton's excitonic content.
(d) The ring peak’s distance from the pump center plotted against the corresponding blueshift. 
This real-space data for each detuning was measured simultaneously with the corresponding k-space data analyzed in Fig.~\ref{S_fig:dispersion_detunings}.
}
\label{S_fig:ring_detunings}
\end{figure*}
\begin{figure}
\centering
\includegraphics[width=0.7\linewidth]{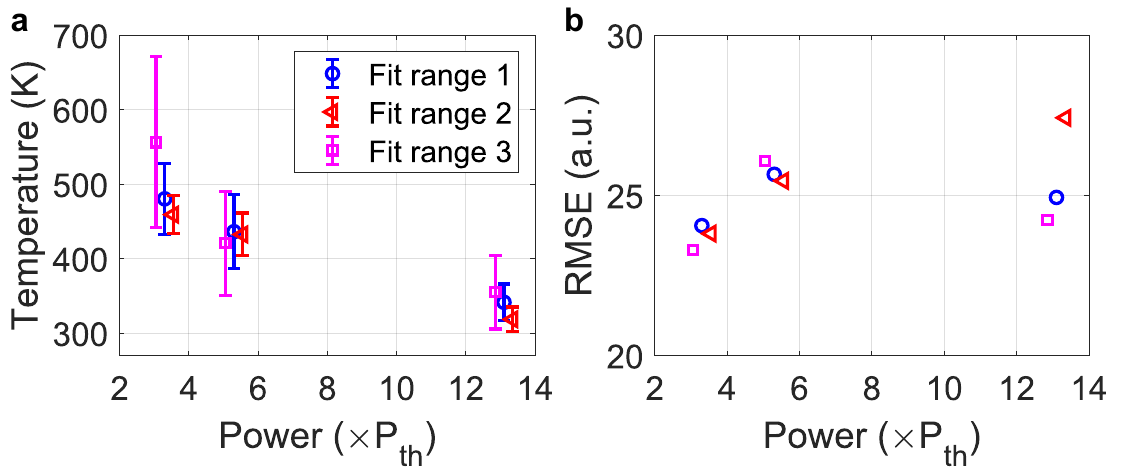}
\caption{Fit sensitivity analysis. (a) Temperatures obtained from Maxwell–Boltzmann fits over different spectral ranges: Fit range 1 (blue circles) corresponds to the 50~meV window used in the manuscript Fig.~\ref{fig:5}; Fit range 2 (red triangles) uses a wider 70~meV window; and Fit range 3 (magenta squares) uses a narrower 30~meV window. The wider and narrower windows are expanded or reduced symmetrically around the fitting range indicated in manuscript Fig.~\ref{fig:5}. The error bars indicate 95\% confidence bounds. (b) Root‑mean‑squared error (RMSE) of each fit shown in (a); the values are reported in the same (arbitrary) intensity units as the spectra in Fig.~\ref{fig:5}(d–f), enabling direct comparison. For clarity, the data points in both panels are offset by $\pm0.25$ units along the x‑axis to resolve overlapping markers.}
\label{S_fig:fit_analysis}
\end{figure}
\begin{figure}
\centering
\includegraphics[width=\linewidth]{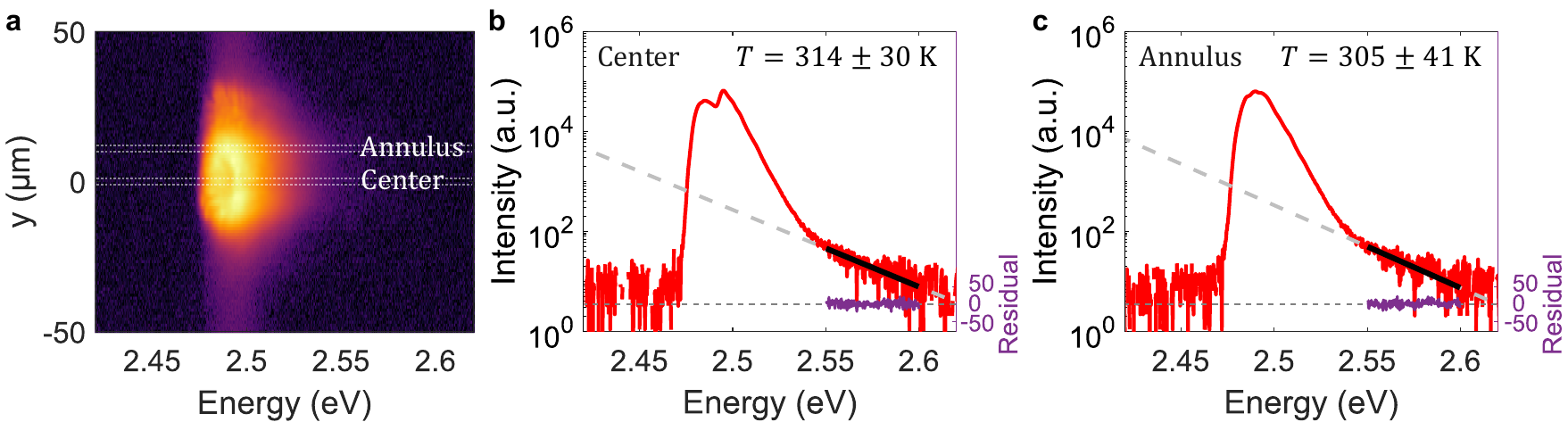}
\caption{Spatial dependence of fit temperature. Comparison of the fitted thermal tails in different spatial regions (center vs. annulus). (a) Real‑space‑resolved emission spectrum for the pump 13.1$P_{\mathrm{th}}$, corresponding to manuscript Fig.~\ref{fig:5}c. (b–c) Fits of the thermal tail to the Maxwell–Boltzmann distribution for the (b) center and (c) annulus regions, with the extracted fit temperatures and their 95\% confidence intervals. Fit residuals are plotted in linear scale on the right‑hand y‑axis. The line spectra in panels (b–c) are obtained by integrating over the respective regions indicated in (a) within a $\pm$3~µm range.}
\label{S_fig:fit_spatial}
\end{figure}

\textbf{Supplementary Figure~\ref{S_fig:DBR_ref}} 
Reflectivity spectrum of the dielectric distributed Bragg reflector (DBR).\\
\textbf{Supplementary Figure~\ref{S_fig:ASE_PD}} 
Power-dependent emission from the neat DPAVB:PVA film, showing the onset of amplified spontaneous emission (ASE).\\
\textbf{Supplementary Figure~\ref{S_fig:more_lasing_info}} 
Power-dependent photoluminescence and linewidth for different combinations of excitation and collection polarizations.\\
\textbf{Supplementary Figure~\ref{S_fig:transient}} 
Time-resolved photoluminescence decay traces of the microcavity emission at various excitation powers.\\
\textbf{Supplementary Figure~\ref{S_fig:Lasing_in_other_det}} 
Power-dependent emission of microcavities at different cavity–exciton detunings.\\
\textbf{Supplementary Figure~\ref{S_fig:ASE_damage}} 
Real-space images of neat DPAVB:PVA films under increasing pump fluence, showing irreversible damage associated with ASE cycling.\\
\textbf{Supplementary Figure~\ref{S_fig:polariton_no_damage}} 
Real-space images of the polariton microcavity under repeated threshold cycling, demonstrating fully reversible radial emission and absence of damage.\\
\textbf{Supplementary Figure~\ref{S_fig:polariton_photo}} 
Photograph of the microcavity sample and spatially resolved PL spectra demonstrating detuning variation due to the spin-coating thickness gradient.\\
\textbf{Supplementary Figure~\ref{S_fig:Transfer matrix}} 
Transfer-matrix simulations of full and empty cavities and measured optical constants of the DPAVB:PS film.\\
\textbf{Supplementary Figure~\ref{S_fig:lasing_reflectivity}} 
Angle-resolved reflection and emission below and above threshold at different pump powers.\\
\textbf{Supplementary Figure~\ref{S_fig:dispersion_detunings}} 
Power-dependent blueshift and angle shift of the emission maximum and their correlation for three detuned cavities.\\
\textbf{Supplementary Figure~\ref{S_fig:ring_detunings}} 
Power-dependent real-space emission profiles, ring-radius analysis, and corresponding blueshift for three detuned cavities.\\
\textbf{Supplementary Figure~\ref{S_fig:fit_analysis}} 
Fit sensitivity analysis.\\
\textbf{Supplementary Figure~\ref{S_fig:fit_spatial}} 
Spatial dependence of fit temperature.\\
\newpage

\newpage

\newpage

\end{document}